\documentclass[numberedappendix,iop,twocolumn,twocolappendix]{emulateapj}
\usepackage{color}
\usepackage{threeparttable}

\slugcomment{Not to appear in Nonlearned J., 45.}

\shorttitle{Spectrum of Crab Giant Radio Pulse}
\shortauthors{Mikami et al.}

\begin{document}

\title{Wide-Band Spectra of Giant Radio Pulses from the Crab Pulsar}
\author{Ryo Mikami, Katsuaki Asano}
\affil{Institute for Cosmic Ray Research, The University of Tokyo, 
    Kashiwa, Chiba 277-8582, Japan}
\email{mikami@icrr.u-tokyo.ac.jp,asanok@icrr.u-tokyo.ac.jp}
\author{Shuta J. Tanaka}
\affil{Department of Physics, Faculty of Science and Engineering,
Konan University, 8-9-1 Okamoto, Kobe, Hyogo, 658-8501, Japan}
\author{Shota Kisaka}
\affil{Department of Physics and Mathematics, Aoyama Gakuin University, Sagamihara, Kanagawa, 252-5258, Japan}
\author{Mamoru Sekido, Kazuhiro Takefuji}
\affil{Kashima Space Technology Center, National Institute of Information and Communications Technology, Kashima, Ibaraki 314-8501, Japan}
\author{Hiroshi Takeuchi}
\affil{Institute of Space and Astronautical Science, Japan Aerospace Exploration Agency, Sagamihara, Kanagawa 252-5210, Japan}
\author{Hiroaki Misawa, Fuminori Tsuchiya}
\affil{Planetary Plasma and Atmospheric Research Center, Tohoku University,
 Sendai, Miyagi 980-8578, Japan }
\author{Hajime Kita}
\affil{Department of Geophysics, Graduate School of Science, Tohoku University,
 Sendai, Miyagi 980-8578, Japan }
\author{Yoshinori Yonekura}
\affil{Center for Astronomy, Ibaraki University, 
 2-1-1 Bunkyo, Mito, Ibaraki 310-8512, Japan}
\and
\author{Toshio Terasawa}
\affil{iTHES Research Group, RIKEN, 
   Wako, Saitama 351-0198, Japan}

\begin{abstract}
We present the results of the simultaneous observation of the giant radio pulses (GRPs)
from the Crab pulsar at 0.3, 1.6, 2.2, 6.7, and 8.4 GHz with four telescopes in Japan.
We obtain 3194 and 272 GRPs occurring at the main pulse and the interpulse phases, respectively. 
A few GRPs detected at both 0.3 and 8.4 GHz are the most wide-band 
samples ever reported.
In the frequency range from 0.3 to 2.2 GHz,
we find that about 70\%  or more of the GRP spectra
are consistent with single power laws 
and the spectral indices of them are distributed from $-4$ to $-1$. 
We also find that a significant number of GRPs have such a hard spectral index (approximately $-1$) that the fluence at 0.3 GHz
is below the detection limit (``dim-hard'' GRPs).
Stacking light curves of such dim-hard GRPs at 0.3 GHz,
we detect consistent enhancement compared to the off-GRP light curve. 
Our samples show apparent correlations between the fluences and the spectral hardness,
which indicates that more energetic GRPs tend to show
softer spectra.
Our comprehensive studies on the GRP spectra are useful materials to verify the GRP model of fast radio bursts in future observations.
\end{abstract}


\section{Introduction}
\label{intro}
From several pulsars,
sporadic intense radio pulses that are much brighter than the normal pulse flux,
called ``Giant Radio Pulses (GRPs)", are observed.
The distinctive feature of GRPs is that their amplitude distribution 
follows a power law \citep[e.g.][]{AG72}, while that of ``normal'' pulses follows a log-normal distribution \citep[][]{BS12}.
GRPs are detected from about 10 pulsars, including some young pulsars such as 
the Crab pulsar (PSR B0531+21) \citep[e.g.][]{Ma11} and PSR B0540-69 \citep{JR03},
and some millisecond pulsars such as PSR B1937+21 \citep{So04} and PSR B1821-24A \citep{Bi15}.
Even some of relatively slowly rotating pulsars, such as PSR B0950+08 \citep[e.g.][]{Ts15},
show GRPs, and in other cases, e.g. PSR B1133+16 \citep[][]{KP15},
GRP-like anomalous intense pulses are detected.
Although several theoretical models of GRPs have been proposed \citep[e.g.][]{Ly07,Pe04}, the emission mechanism of GRPs has not yet been revealed.

The GRPs from the Crab pulsar are the most extensively observed,
and they have provided a lot of information.
The Crab GRPs were detected in the frequency range from 20 MHz \citep[][]{El13} to 46 GHz \citep[][]{Ha15}.  
The Crab GRPs occur at the two rotational phases:
the main pulse or the interpulse phases of the normal pulse \citep[][]{Co04}.
The GRP light curves often consist of a number of (sub-)nanosecond bursts,
so-called nanoshots. Their peak flux densities sometimes exceed
several million Jy, and their brightness temperatures
sometimes reach $\sim10^{41}$K \citep{HE07}, which are higher than any other astronomical phenomena ever found \citep[e.g. Figure 7 of][]{Ao14}.

From simultaneous multi-frequency observations,
the spectra of the Crab GRPs have been investigated in several studies.
\citet[][]{SB99} showed that the spectral indices distribute
between $-4.9$ and $-2.2$ for the GRPs simultaneously detected at 0.6 and 1.4 GHz.
\citet[][]{Or15} also reported the index distribution between $-4.9$ and $-3.6$ based on observations at 0.2 and 1.4 GHz.
At 1.30--1.45 GHz, \citet[][]{Ka10} claimed that the indices of the GRPs occurring at the main pulse and the interpulse phases were $-1.4\pm3.3$ and $-0.6\pm3.5$, respectively.
They also reported that some fractions of GRPs were not detected at the frequency sub-bands of 1.30--1.45 GHz, indicating narrow-band spectral feature intrinsic to the GRPs.
From simultaneous observations at 0.6, 1.65, and 4.85 GHz,
\citet[][]{Po08} claimed that the average GRP spectra
seem to be bent harder with increasing frequency.
Even in \citet[][]{Po09}, based on 27 samples observed at 0.6, 1.4, and 2.2 GHz,
the spectral hardening with frequency was claimed again.

Recently, the discovery of another sporadic phenomenon, fast radio bursts
(FRBs), has been reported \citep[e.g.][]{Lo07,tho13}.
FRBs are radio transients that have durations of a few milliseconds and have high
dispersion measures (hereafter DMs) that exceed that expected from the interstellar medium
of our own Galaxy.
Considering the large DM,
the origin of the FRBs is extragalactic, but the source object itself is not unknown.
Repeating bursts were observed from a position of FRB 121102 \citep[][]{Sp16,Sc16} with a power-law distribution
of its fluence \citep[][]{WY16,LK16}, which rules out catastrophic events such as a
neutron star merger \citep[e.g.][]{To13} and a white dwarf merger \citep[][]{Ka13} at least for this FRB.
GRPs from young pulsars are discussed
as one of the candidates of the FRB \citep[e.g.][]{CW15,Co15}.
Revealing the spectral features of known GRPs are indispensable to test the young pulsar model as a candidate of FRBs,
and provide the expected event rate in other energy bands \citep[e.g.][]{Be16}.

Motivated by the situation mentioned above, we operated simultaneous multi-frequency observations of the Crab GRPs
at 0.3, 1.6, 2.2, 6.7, and 8.4 GHz with the four radio telescopes in Japan,
and report the results in this paper.
In most of the previous studies, the sample number of GRPs is
insufficient to discuss the statistical properties of the spectra,
except for \citet[][]{Po08} and \citet[][]{Ka10}.
In \citet[][]{Po08} focused only the average spectrum based on $\sim 1500$ samples,
while the frequency separation ($\Delta \nu/\nu \sim 0.1$) in the observation of \citet[][]{Ka10}
is too narrow to discuss the global structure of the spectra.
Our study is carried out in order to overcome those weak points.
The outline of our observation and data analysis are described in Section \ref{obs}.
Spectral analyses with about 3200 main pulse GRPs and about 250 interpulse GRPs 
at a frequency range from 0.3 to 2.2 GHz, are shown in Section \ref{3band}.
We also show some examples that apparently contradict with a single power law.
Wider spectra from 0.3 to 8.4 GHz are shown in Section \ref{5band}.
Our results are summarized with a discussion in Section \ref{dis}.

\section{Observation and Analysis}
\label{obs}
Our radio wide-band observational campaign was conducted
from UTC 14$^{h}$30$^{m}$ on September 6 to 04$^{h}$00$^{m}$ on 2014 September 7.
Simultaneous observations at all observatories were overlapped
from UTC 21$^{h}$35$^{m}$ on September 6 to 02$^{h}$20$^{m}$ on September 7.
We summarize our observation in Table \ref{t1}.

\begin{table*}[!htb]
\begin{center}
\caption{Observation summary. \label{t1}}
{\scriptsize
\begin{tabular}{cclccccc}
\hline 
   \begin{tabular}{c} Observatory \\(Longitude, Latitude)\\ Antenna \end{tabular}
 & \begin{tabular}{c} Start Time$^{\rm a}$ \\ (Duration)  \end{tabular}
 & \begin{tabular}{c} Center Frequency \\ $\nu_{\rm c}$        [MHz] \\ (Band Name)$^{\rm b}$ \end{tabular}
 & \begin{tabular}{c} Bandwidth$^{\rm c}$        \\ $\Delta \nu$ [MHz] \\ (Bits/Sample)  \\ Polarization$^{\rm d}$ \end{tabular}
 & \begin{tabular}{c} SEFD$^{\rm e}$      \\ (Jy)      \end{tabular}
 & \begin{tabular}{c} $S_{\rm CN}^{\rm f}$    \\ (Jy)      \end{tabular}
 & \begin{tabular}{c} Selection    \\ Threshold \\ ($\Delta t, \mbox{S/N}$)     \end{tabular}
\\ 
\hline
   \begin{tabular}{c} Iitate \\ (140$^{\rm o}$ 40'E, 37$^{\rm o}$ 42'N)\\31m$\times$16.5m, \\ 2 asymmetric \\ offset dishes  \end{tabular}              
 & \begin{tabular}{c} 15:11 \\ (680 minutes) \end{tabular} 
 & ~~~~~325.1 (P)~~~& 4(8) Linear$^{\rm g}$
 & 1066-1326  & 1164$\pm$128 & $500\mu$s, $5\sigma$
\\
\hline
\begin{tabular}{c}Kashima \\ (140$^{\rm o}$ 40'E, 35$^{\rm o}$ 57'N)\\34m dish \end{tabular}
& \begin{tabular}{c} 14:30 \\ (810 minutes) \end{tabular} 
& 
    \begin{tabular}{l} ~~1586(L2) ~~~~$\equiv$ LL \\ 
      $
        \left .
          \begin{array}{c} 1674(\rm L6) \\ \underline{1696}(\rm L7) \\ 1718(\rm L8) \end{array}
        \right \} \equiv {\rm LH}
      $
  \end{tabular}
& \begin{tabular}{r}     32(4) R ~~~~~~~   \\ 
      $ \left .
        \begin{array}{c} 32 (4) ~\rm R    \\     32 (4) ~\rm R    \\    32 (4) ~\rm R  \end{array}
        \right \} \underline{76}
      $
  \end{tabular}
& \begin{tabular}{r} 422-543 \\ 357-443 \\ 400-549 \\ 416-525 \end{tabular}
& \begin{tabular}{r} 816$\pm$16\\ 801$\pm$16\\ 797$\pm$16\\ 794$\pm$16\end{tabular}
& \begin{tabular}{r} $10\mu$s, $6\sigma$\\ \  \\ $10\mu$s, $6\sigma$\\ \ \end{tabular}
\\ 
\hline
 \begin{tabular}{c} Usuda \\(138$^{\rm o}$ 22'E, 36$^{\rm o}$ 08'N)\\ 64m dish \end{tabular}
& \begin{tabular}{c} 20:50 \\ (430 minutes) \end{tabular} 
& \begin{tabular}{l} 
     $ \left .
       \begin{array}{c} 2198 (\rm S1) \\ 2224 (\rm S2) \\ \underline{2250} (\rm S3) \\ 2276 (\rm S4) \\ 2302 (\rm S5) \end{array}
       \right \} \equiv {\rm S}
     $
   \\
    ~~8445 (X) 
  \end{tabular}
& \begin{tabular}{r} 
       $ \left . \begin{array}{c}  32(4)\rm R    \\     32(4)\rm R    \\     32(4)\rm R    \\    32(4)\rm R  \\     32(4)\rm R \end{array} \right \} 
          \underline{136} $
      \\    32(4) R ~~~~~~~
  \end{tabular}
& \begin{tabular}{r} 112-122   \\ 106-116 \\ 103-112 \\ 102-111 \\ 107-117 \\  100-110 \end{tabular}
& \begin{tabular}{r} 661$\pm$13\\ 657$\pm$13\\ 653$\pm$13\\ 648$\pm$13 \\ 644$\pm$13 \\ 125$\pm$15\end{tabular}
& \begin{tabular}{r} \    \\ \  \\  $10\mu$s, $6\sigma$ \\ \  \\ \  \\  $1\mu$s, $9\sigma$ \end{tabular}
\\
\hline
 \begin{tabular}{c} Takahagi \\ (140$^{\rm o}$ 42'E, 36$^{\rm o}$ 42'N)\\32m dish \end{tabular}
 & \begin{tabular}{c} 15:00 \\ (730 minutes) \end{tabular} 
 & ~~~~6672 (C) & 16(4) L ~~~~
 &  289-359  & 406$\pm$9 & $1\mu$s, $10\sigma$
\\ \hline
\end{tabular}
}
\noindent
\end{center}
{\scriptsize 
$^{\rm a}$UTC on 2014 September 6. \\
$^{\rm b}$The channel L2 is renamed LL. The channels L6--L8 and S1--S5 are synthesized, and renamed LH and S, respectively (see section \ref{GRPsel}).
The central frequencies after the synthesizing procedure are underlined.
\\
$^{\rm c}$Effective bandwidth after the synthesizing procedure are underlined. \\
$^{\rm d}$R/L: Right/left-handed circular polarization. \\
$^{\rm e}$System equivalent flux density. Characteristic values during our observing session are shown. \\
$^{\rm f}$Received flux density of the Crab nebula. See Section \ref{binning} and Appendix \ref{est}.   \\
$^{\rm g}$Horizontal component.
}
\end{table*}

The observation at 0.3 GHz (P band) was made with the Iitate Planetary Radio Telescope (IPRT),
Tohoku University \citep[][]{Mi03,Ts10}.
The telescope consists of two sets of asymmetric offset dish antennae of 31m$\times$16.5m. 
To record the data, we used the K5/VSSP32 sampler \citep[][]{Ko03}.
The signal was 8-bit sampled at the Nyquist rate of 8 MHz with 4 MHz bandwidth.
Only the horizontal polarization signal was recorded. 

The GRP data for 1.4--1.6 GHz (L band) was obtained with
the 34m telescope at the Kashima Space Technology Center \citep{Tak16} operated
by the National Institute of Information and Communications Technology (NICT). 
We recorded the data with the ADS3000+ recorder \citep[][]{Ta10}.
The signals were 4-bit sampled at the Nyquist rate of 64 MHz,
and the data were divided into eight channels (L1--L8) with 32 MHz bandwidths.
Only the right-handed circular polarization (RHCP) signal was received
at all the frequency channels.
At Kashima observatory, to avoid radio frequency interference (RFI)
from cell phone base stations, a superconductor filter is installed.
Nevertheless, frequent RFI occurred during our observation at some frequency channels.
We only use the four channels of L2, L6, L7, and L8, which are less affected by RFI in subsequent analysis.

The 64m telescope, belonging to the Usuda Deep Space Center
of the Institute of Space and Astronautical Science (ISAS),
took part in the observation at 2.2 GHz (S band) and 8.4 GHz (X band).
The data acquisition system is the same as that of the Kashima observatory.
We recorded the data with the ADS3000+ recorder. 
The signals were four-bit sampled at the Nyquist rate of 64 MHz 
and the data were divided into eight channels with 32 MHz bandwidths.
Seven of their channels were assigned to S-band data (S1--S7),
and the other one channel was used for X band data.
Only the RHCP signal was received at all the frequency channels.
Since the receiver sensitivities in S6 and S7 were insufficient,
only the data at S1--S5 and X bands were analyzed. 

The observation at 6.7 GHz (C band) was made with the Takahagi 32m antenna, which is a branch of Mizusawa VLBI observatory of National Astronomical Observatory of Japan \citep{Yo16}. We recorded the data with the K5/VSSP32 sampler. The signals were four-bit sampled at the Nyquist rate of 32 MHz with 16 MHz bandwidth. Only the LHCP signals was received.

During the simultaneous observation period,
calibration observations were intermittently made in all of the observatories described above.
At L band in Kashima, the data for the periods of RFI were manually removed.
As a result, the effective simultaneous observation time becomes 2.9 hr.

We determined the DM by aligning the peak times of some bright main pulse GRPs
at the frequency channels of L band (L2, L6, L7, and L8) and S band (from S1 to S5).
The times of arrival (TOAs) of the GRPs were converted to Barycentric Dynamical Times at the solar system barycenter
with the pulsar timing package TEMPO2 \citep{Ho06}.
The accuracy of the TOAs in our analysis is within $1\ \mu$s.
According to the dispersion delay formula between frequencies $\nu_1$ and $\nu_2$,
\begin{equation}
\label{dispersion}
\Delta t_{12} \simeq \frac{e^{2}}{2\pi m_{e}c}\left(\frac{1}{\nu_{1}^{2}}-\frac{1}{\nu_{2}^{2}}\right)
\mbox{DM},
\end{equation}
the DM value was determined 
as 56.764 pc/cm$^{3}$ in our observation period\footnote{This value is roughly consistent with those reported by Jodrell bank Crab pulsar monthly ephemeris (\citealt{Ly93}, http://www.jb.man.ac.uk/pulsar/crab.html), which are $56.769\pm0.005$
pc/cm$^{3}$ and $56.772\pm0.005$ pc/cm$^{3}$ on 2014 August 15, and 2014 September 15, respectively. Furthermore, taking account of intra-month variation of DM reported \citep[e.g.][]{Ku08}, the difference between our estimated value and the monthly reported value is consistent.}. 
The sampled data for each frequency channel were dedispersed with this DM value,
based on the coherent dedispersion method \citep[][]{HR75,LK04}. 

\begin{figure}[!htb]
\epsscale{1}
\plotone{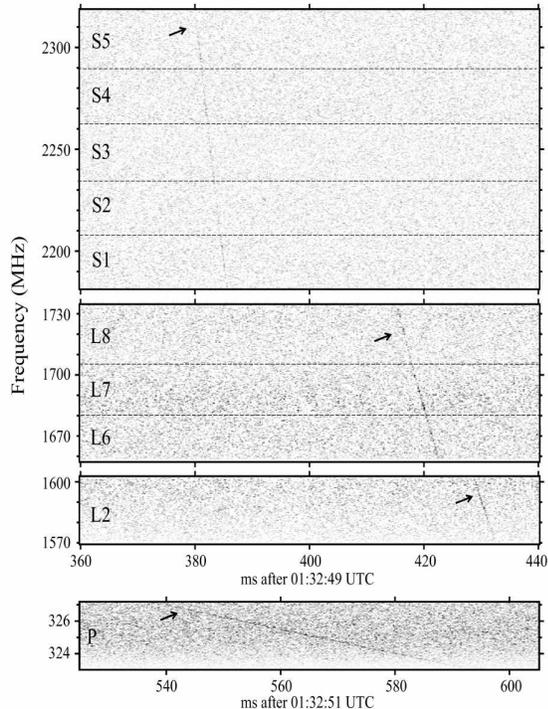}
\figcaption{Dynamic spectra of the GRP \#2677, which occurred at 
01:32:49-51 UTC on 2014 September 7.  This GRP occurred at the main pulse phase.\label{dynsp1}}
\end{figure}

\begin{figure}[!htb]
\epsscale{0.8}
\plotone{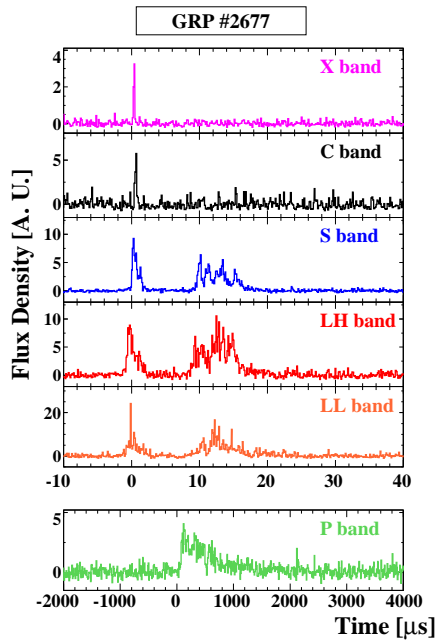}
\figcaption{Dedispersed light curves of the GRP \#2677
(same GRP as that shown in Figure \ref{dynsp1}),
which is simultaneously detected at all the bands.
The curve in the P band is smoothed with $10\ \mu$s time-bin.
The other curves are smoothed with $125$ ns.
The spectrum of this GRP is shown in Figure \ref{golden}.\label{sp1}}
\end{figure}

Examples of the pre-dedispersed dynamic spectra and the dedispersed pulse profiles of a GRP
are shown in Figures \ref{dynsp1} and \ref{sp1}, respectively.
In Figure \ref{dynsp1}, the dispersive delays of the GRP signal
expected from equation (\ref{dispersion}) are clearly shown, indicating that this signal 
is a real GRP.
The GRP profile shown in Figure \ref{sp1} was simultaneously observed with the frequency coverage over $\sim1.5$ decades, which is the most wide-band ever reported.

\subsection{Systematic search of GRPs}
\label{GRPsel}
Below, we describe our methods of GRP search.
To extract real GRPs, we set two criteria for the GRP selection:
(1) a peak signal-to-noise ratio (S/N) value of dedispersed data exceeds the threshold S/N tabulated in Table \ref{t1},
and (2) the signal appears at the rotational phase of the main pulse or interpulse.

We smoothed the time series of the dedispersed data for each channel
with time resolutions $\Delta t$ tabulated in Table \ref{t1},
which are adjusted to draw the typical GRP light curve with a few time-bins.
From off-pulse phases, we obtained the noise flux,
which is mainly due to the Crab nebula background.
Based on the noise level and fluctuation, we calculated the S/N
of the time series of the data.
In order to improve the S/N and mitigate the effects of
the interstellar scintillation (see Section \ref{scintillation}),
the data from L6 to L8 channels, and the data from S1 to S5 channels were
merged taking into account
the overlaps of the frequency intervals of those channels.
Hereafter, we call those merged frequency bands ``LH band'' and ``S band'', respectively.
We also treat the L2 (hereafter, ``LL'') band separately from the LH band,
because the frequency separation between the L2 and the L6 bands is larger than their bandwidths.

\begin{figure}[!h]
\epsscale{1}
\plotone{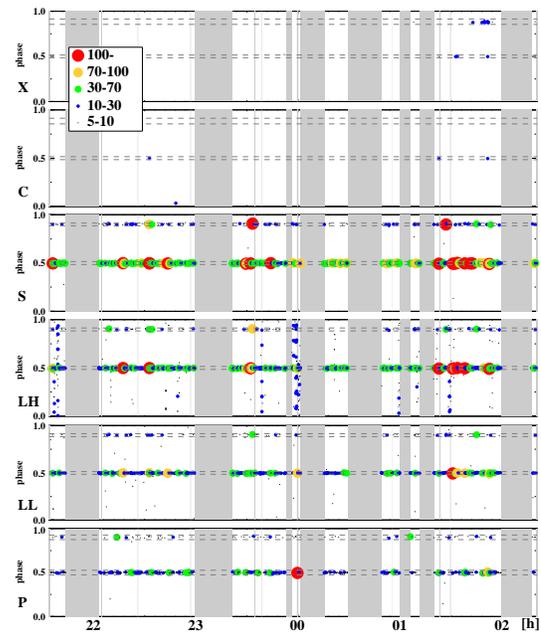}
\caption{Scatter plots of phase and time for the GRP candidates from 21$^{h}$30$^{m}$ on 2014
September 6th to 02$^{h}$20$^{m}$ on 2014 September 7th
for all the frequency bands. The gray shaded areas are the excluded periods:
the calibration observation or the period heavily affected by RFI. The GRP selection phases are denoted by the horizontal dashed lines.
}
\label{candidate}
\end{figure}

RFI and statistical fluctuations of the noise give rise to fake signals.
The thresholds S/N for the GRP selection (hereafter, ``selection thresholds'')
were determined to
collect as many GRPs as possible and to avoid contamination of fake GRPs.
The GRP candidates, whose peak S/N values exceed the selection thresholds, are plotted
in Figure \ref{candidate}, where the peak time of the main pulse in LL--S bands
is defined as 0.5 in the rotational phase. 
For P, LL, LH, and S bands,
the phase ranges defined to accept signals as GRPs
were empirically determined from the distribution of the GRP candidates in our samples.
For LL, LH, and S bands, the phase widths are 0.03 rotational period
centering at the peak phases
for both the main pulse and interpulse phase ranges.
The width for the P band is a 0.05 rotational period.
At C and X bands, the number of the candidates is too small to determine
the phase ranges for the GRP selection.
We adopted the same phase ranges as those at S band for the main pulse.
For the interpulse phase, we shift the central position 0.015 rotational period
earlier than that in S band with a width of 0.06 rotational period
following the results of \citet{Co04}, who reported that
the occurrence phases of GRPs shift $\sim 0.03$ period earlier at above $\sim 4$ GHz.
Even in the other phase ranges so-called HFC1/2, 
\citet{Je05} and \citet{Mi12} reported the detections of GRPs at X band 
\citep[however, see][]{Co04}.
Though the HFC GRPs are worth investigating, the wide phase range
and its low occurrence rate make it difficult to distinguish real GRPs 
from fake signals due to RFI in our samples.
We do not discuss HFC GRPs in this paper.

Some candidates appearing outside the defined phases are identified as fake signals.
We estimate the expected number of the fake signals appearing accidentally at the GRP selection phases.
The expected number of such contamination is at most $\sim 5.2\pm2.3$ (at the LH band,
the errors are estimated assuming that the fake signals follow the Poisson statistics.)
during the entire time of the simultaneous observation at all the observatories
(see Table \ref{contami}).
The contamination is almost negligible compared to our sample number ($\sim 3000$).
The fake signal numbers for C and X bands are very sensitive to the threshold S/N,
so that we need to set a relatively higher S/N for the selection threshold in those bands
as shown in Table \ref{t1}.

\begin{table}[!htb]
\begin{center}
\caption{Expected number of fake signals appearing accidentally at the GRP selection phases during entire time of the simultaneous observation.
 \label{contami}}
\begin{tabular}{cc}
\hline
Frequency & Number   \\
Band &  \\
\hline
X   &  $9.9\times10^{-2}$ \\
C  &  $9.9\times10^{-2}$ \\
S    &  $0.45$ \\
LH &  $5.2$ \\
LL &  $2.5$ \\
P &  $1.0$ \\
\hline
\end{tabular}
\end{center}
\end{table}

As a result, we obtained 3194 GRPs occurring at the main pulse phase (hereafter MPGRPs),
and 272 GRPs occurring at the interpulse phase (hereafter IPGRPs)
during the effective observation time of 2.9 hr.
Only 5 MPGRPs and 19 IPGRPs exceeds the selection thresholds at C and/or X band(s).
All of those 5 MPGRPs were simultaneously detected at lower frequency bands (P--S).
In contrast, all the 19 IPGRPs were detected only at X band.

\subsection{Scintillation Effects}
\label{scintillation}
Extrapolating the data in Figure 3 of \citet[][]{RL90},
the timescales of the refractive interstellar scintillation (RISS)
are estimated to be about 12 hr or longer for the frequency range lower than $\sim 2$ GHz.
Therefore, at these frequency bands, the intensity modulation
due to the RISS can be almost negligible during our about 5 hr simultaneous observing
session.
In contrast, the RISS timescales for C and X band observations
are about one hour so that the pulse flux could be
modulated during our observing session.
Indeed, in the 8.9 GHz observation of the Crab pulsar,
\citet[][]{Bi11} showed that the numbers of the detected GRPs
during two observing sessions of 2 hr in the same day
were quite different.
This variation of the detection rate may be due to the RISS rather than
the intrinsic variation of the GRP activity.
In our observation at C or X band, a cluster of GRP candidates can be also seen around 01$^{h}$30$^{m}$
on September 7 (see Figure \ref{candidate}). Therefore, the fluences at our C and X bands may be
affected by the RISS.

As for the diffractive interstellar scintillation (DISS),
\citet[][]{Co04} measured the scintillation bandwidths as
$<0.024$ MHz, $<0.8$ MHz, and $2.3\pm0.4$ MHz
at 0.43, 1.48, and 2.33 GHz, respectively.
In our observation at P, L, and S bands, they are at least an order of magnitude smaller
than our bandwidths of the frequency channels (see Table \ref{t1}).
The scintillation modulation is presumed to be averaged out at those frequency bands
\footnote{Note that we obtained some GRPs that show spectral structures
in narrow frequency ranges, as shown in Appendix \ref{binning1}.
This may indicate that the DISS effect has not been completely understood yet.
Alternatively, those spectral structures may be intrinsic in GRPs.}.
Also, at L and S bands, the synthesis procedure of the multiple channels
described in Section \ref{GRPsel} may reduce the effect of the DISS.
In contrast, extrapolating the result in \citet[][]{Co04}
with $\nu^{4.4}$ scaling (corresponding to the Kolmogorov spectrum
for the interstellar electron density fluctuation),
the scintillation bandwidths for C and X bands
are larger than the observation bandwidths.
Therefore, the fluences at our C and X bands may be
significantly affected by the DISS.
Indeed, in the observation by \citet[][]{Co04}, 
the mean pulse flux density was deeply modulated on timescales as short as five minutes
at frequencies $\gtrsim 3$ GHz.

As we described above, the fluences at P--S bands may be less affected by the interstellar scintillation so that the fluences
in this frequency range are relatively reliable compared to the fluences at the higher frequency bands.
Therefore, we first discuss the spectra at P--S bands in Section \ref{3band},
and then show the spectra including the fluences at C and X bands in Section \ref{5band}.

\subsection{Fluence Estimate}
\label{binning}

Hereafter, we focus on the time-integrated flux (fluence) of the GRPs
rather than the peak flux,
because the scatter broadening of the pulse profile
depends on frequency \citep[e.g.][]{Cr70}.
In our samples, the GRP durations at LL--X bands distribute widely,
while the typical one is $\sim2\ \mu$s.
The P-band pulse profiles show a typically $\sim1$ ms extent.
When we estimate the fluences,
we generate time-series of the S/N values with 1 $\mu$s time bin
for LL--X bands, and $500 \mu$s time bin for P band.
Then, the S/N values of all the frequency bands are transformed to
the flux density calibrated by reference sources such as Cas A, Cyg A, and so on.
In the calibration procedure, we also estimate a received fraction of the flux density
from the background nebula taking into account the sizes of the antenna beam and the nebula,
and the spatial intensity profile of the nebula.
The details of the calibration procedure are described in Appendix \ref{est}.
In the time integration procedure to obtain the fluence $F$,
a too long time interval compared to the actual duration of a GRP
leads to an excessively large error.
We set the time interval individually for each GRP
to contain the entire pulse sufficiently.
The details to determine the time interval are summarized in Appendix \ref{interval}.
Since the sample numbers in C and X bands are small, we do not establish a systematic way
to determine the time interval for those bands.
The intervals determined manually will be explicitly shown in the light curves in Section \ref{5band}.
As long as a GRP is selected in a certain band with the method described in Section \ref{GRPsel},
the fluences or their upper-limits are estimated for all the bands.

\begin{figure}[!htb]
\epsscale{1.2}
\plotone{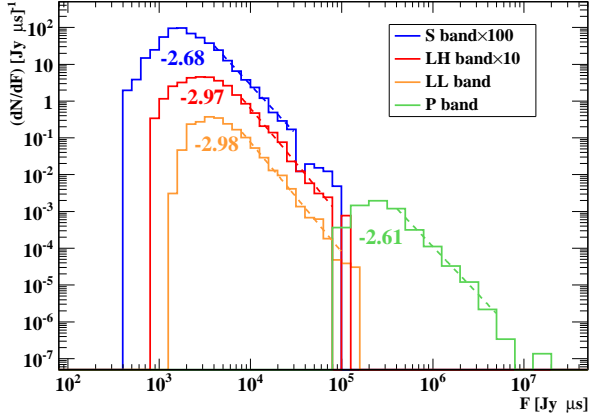}
\figcaption{
\footnotesize
Fluence histograms $dN/dF$ of MPGRPs in the four bands P, LL, LH, and S.
Dashed lines show the fitted power-law function $\propto F^{\Gamma}$.
The numbers in the figure are corresponding $\Gamma$.
\label{fig:fluenceDistribution}}
\end{figure}

Figure \ref{fig:fluenceDistribution} shows
the fluence histograms $dN/dF$ of MPGRPs for the P, LL, LH, and S bands.
Representing the fluence distribution by a power law
$dN/dF \propto F^\Gamma$ for the high-fluence portion,
the index $\Gamma$
obtained at each band is shown in Table \ref{tFluenceIndex}.
The obtained indices $\Gamma$ are consistent with the previous studies
\citep[e.g.][]{Mi12}.

\begin{table}
\begin{center}
\caption{Power-law index $\Gamma$ for the fluence distribution.}
\label{tFluenceIndex}
\begin{tabular}{crr}
\hline
Band & \multicolumn{2}{c}{$\Gamma$ (Sample Number)}\\
\cline{2-3}
 & MPGRP & IPGRP\\
\hline\hline
       P          &  $-2.61^{+0.13}_{-0.15}$ (760)   & $-2.73^{+0.55}_{-0.83}$ (101) \\
LL      &  $-2.98^{+0.11}_{-0.12}$ (2166)  & $-2.88^{+0.37}_{-0.50}$ (171) \\
LH      &  $-2.97^{+0.13}_{-0.15}$  (2574) & $-2.75^{+0.72}_{-0.87}$ (192) \\
S         &  $-2.68^{+0.11}_{-0.13}$ (2681) & $-3.62^{+0.55}_{-0.70}$ (192)\\
\hline
\end{tabular}
\end{center}
\end{table}

\section{GRP Spectra from P to S Bands}
\label{3band}
\begin{figure}[!htb]
\epsscale{1.2}
\plotone{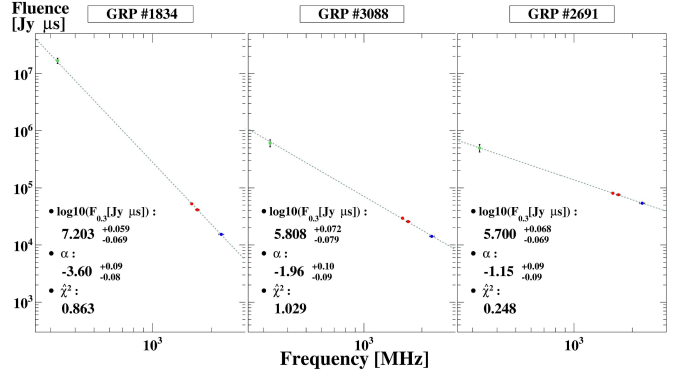}
\figcaption{Examples of the GRP spectra consistent with SPLs.
The best-fit power-law functions are plotted with the dotted lines.
The obtained parameters with 68\% confidence intervals
and the minimum $\hat{\chi}^{2}$
are also shown.\label{sp4}}
\end{figure}

\begin{figure}[!htb]
\epsscale{1.2}
\plotone{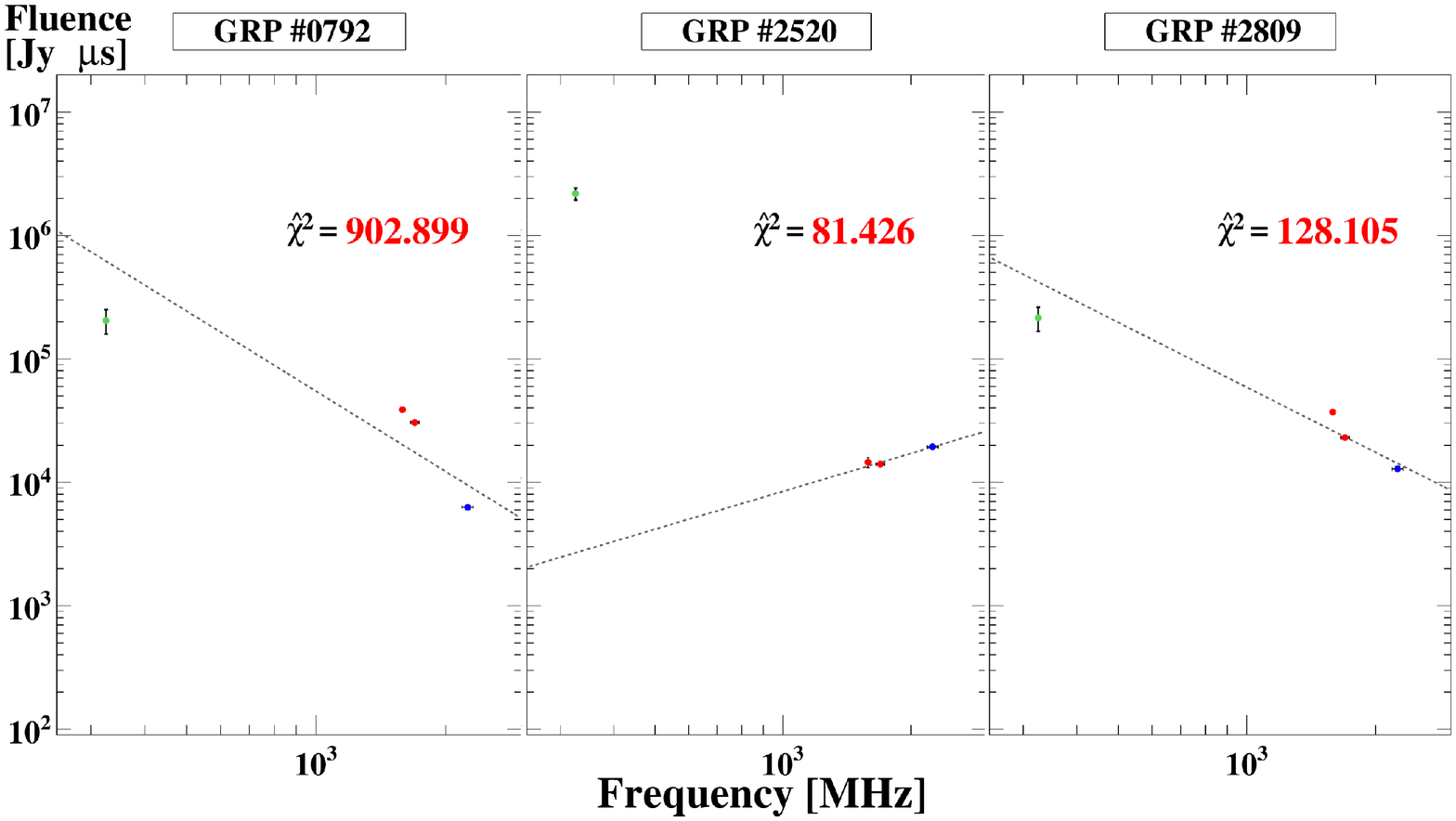}
\figcaption{Examples of the GRP spectra inconsistent with SPLs:
hard-to-soft (left), soft-to-hard (middle),
and other (right) spectra. 
The minimum $\hat{\chi}^{2}$ values and corresponding SPL functions (dotted lines)
are shown.
 \label{chisqhigh}}
\end{figure}

In Figures \ref{sp4} and \ref{chisqhigh}, we show some examples of the spectra. 
Here, we examine what fraction of the wide band spectra in our samples are consistent with single power laws
(hereafter SPLs).
The GRP spectra are fitted by the SPL as a function of frequency $\nu$,
\begin{equation}
\label{fit0.3}
F(\nu)=F_{0.3} \left(\frac{\nu}{325.1\mathrm{MHz}}\right)^{\alpha},
\end{equation}
with a normalization parameter $F_{0.3}$ and index $\alpha$.
The goodness-of-fit test is done by means of the $\chi^{2}$ statistic.
All the fluences from P to S bands are not always determined.
Even if upper limits are included in some frequency bands,
we can test the consistency with a SPL for each GRP by the modified $\chi^{2}$ statistic as
\begin{eqnarray}
\label{modifiedchi}
\hat{\chi}^{2}&=&\sum_i
\chi_{i,{\rm det}}^{2} +\sum_j \chi_{j,{\rm UL}}^{2}.
\end{eqnarray}
For the band $i$, where a certain fluence $F_{i}$ is determined,
its contribution to the $\chi^{2}$ statistic is written as
\begin{eqnarray}
\chi_{i,{\rm det}}^{2}&\equiv&\left(\frac{F_{i}-F(\nu_{i})}{\sigma_{i,{\rm tot}}}\right)^{2},
\end{eqnarray}
where $\sigma_{i,{\rm tot}}$ is the $1\sigma$ total error
for the given time interval in the band $i$ (see Appendix \ref{est}).
For the band $j$, where a fluence upper limit
$F_{j,{\rm max}}$ (see Appendix \ref{est}) is set, its contribution is
\begin{eqnarray}
\chi_{j,{\rm UL}}^{2}&\equiv&-2\ln\int_{-\infty}^{F_{j,{\rm max}}}
\frac{\exp \left( \frac{-\left(F'-F(\nu_{j})\right)^{2}}{2\sigma_{j,{\rm tot}}^{2}}\right)}
{\sqrt{2\pi \sigma_{j,{\rm tot}}^{2}}}
dF',
\end{eqnarray}
\citep[][]{Av80,Sa12}.


We set a critical value of $\hat{\chi}^{2}$ to reject the SPL hypothesis
assuming that $\hat{\chi}^{2}$ follows a $\chi^{2}$ distribution
with $n-2$ degrees of freedom (DoFs) for $n$ data points.
In the case of the spectra for P--S bands,
the number of the data points implies $2$ DoFs for $\chi^{2}$ distribution.
We adopt a critical value of $\hat{\chi}^{2}=5.99$ (significance level of 5\%).

For the GRPs whose spectra are consistent with SPLs,
we estimate the confidence intervals or the upper-limits of
the fitting parameters as follows.
For each pair of the parameters $(F_{0.3}, \alpha)$,
we calculate $\hat{\chi}^{2}$, and express
\begin{equation}
\hat{\chi}^{2}=\hat{\chi}^{2}_{\rm{min}}+\Delta \hat{\chi}^{2},
\end{equation} 
where $\hat{\chi}^{2}_{\rm{min}}$ is the minimum value of $\hat{\chi}^{2}$. 
According to \citet{La76}, we assume that $\Delta \hat{\chi}^{2}$
follows a chi-square distribution with $p$ DoFs,
where $p$ is the number of fitting parameters.
A 68\% confidence interval of
each fitting parameter is that satisfying
\begin{equation}
\Delta \hat{\chi}^{2}=2.3,
\end{equation}
for $p=2$ in our case.

First, we focus on
the GRPs detected at all the four frequency bands.
The fractions of such ideal samples are relatively small,
8.4\% and 18\% for MPGRPs and IPGRPs, respectively (see Table \ref{tcatMP}).
In those samples, we find that 86 of 268 (32\%) MPGRPs, and 27 of 46 (59\%) IPGRPs
are consistent with SPL spectra at a significance level of $5\%$.
The spectral index widely distributes
between $\sim-4$ and $\sim-1$ (see Figures \ref{sp4} and \ref{alphahist-fludiv}), while the spectral index of the normal main pulse and interpulse
are $-3.0$ and $-4.1$, respectively \citep{MH99}.
GRP spectra may show a wider variety in their indices than those of normal pulses.
If the significance level is set to 0.1\% ($\hat{\chi}^2<13.82$),
143 (53\%) MPGRPs, and 35 (76\%) IPGRPs are consistent with SPL spectra.
Furthermore, the goodness of fit depends on the method of the error estimate.
When we impose more conservative errors on the fluences (Method B, see Appendix \ref{binning1}),
the fraction of the consistent spectra with SPLs becomes $\sim90\%$.
Even in the ideal samples, we do not find any strong evidence that the majority of the GRP spectra have
a spectral break or multiple components.

\subsection{Detection Category}
\label{cate} 
In most of our samples, the fluence data points in the given four bands include several upper-limits.
As described below, we categorize our GRPs into the three groups
according to the detection/non-detection in the individual bands.
Since we focus on the wide-band spectra,
we do not discriminate LL and LH bands in the detection category.
The GRPs detected at LL and/or LH bands are categorized as GRPs detected at L band.

\begin{figure}[!htb]
\epsscale{0.8}
\plotone{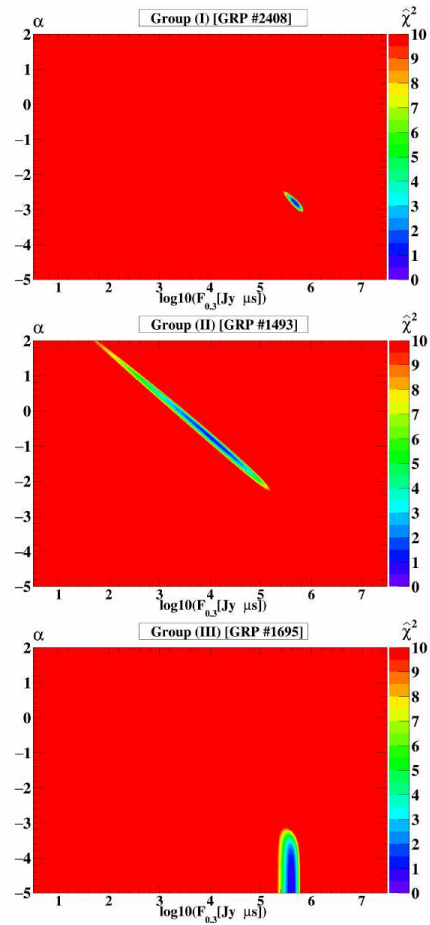}
\figcaption{Behavior of $\hat{\chi}^{2}$ in the parameter space ($F_{0.3}, \alpha$)
for the typical GRPs in Groups (I)--(III) (see Section \ref{cate}). \label{Chimap}}
\end{figure}

The GRPs in Group (I) are detected at multiple frequency bands
including P band.
For Group (I) GRPs, the parameters $F_{0.3}$ and $\alpha$ are determined more precisely than the GRPs in the other groups.
An example of the $\hat{\chi}^{2}$-map of the GRPs in Group (I) is shown
in the top panel of Figure \ref{Chimap}.

Group (II) is defined as
the GRPs detected at both L and S bands,
but not detected (set an upper-limit) at the P band.
For those GRPs, the P-band fluence is estimated by extrapolation using the fluences
at L and S bands. For the samples consistent with SPLs in Group (II),
of course, the extrapolated value does not conflict with the upper-limit
at the P band.
As shown in the middle panel of Figure \ref{Chimap}, the typical allowed region in the parameter space
for Group (II) GRPs is obliquely elongated.
In this case, the spectral index is loosely determined
because of the relatively narrow frequency separation between L and S bands.

The GRPs detected at P band, but neither L nor S band, belong to
Group (III).
For those GRPs, while the fluence at P band, $F_{0.3}$, is precisely determined,
the upper-limits at higher frequency bands merely provide an upper-limit on
$\alpha$.
The $\hat{\chi}^{2}$-map in the bottom panel of Figure \ref{Chimap}
shows the allowed region is vertically elongated.
In this case, we set $\hat{\chi}_{\rm{min}}^{2}=0$.  
Then, we set the 68\% upper limit of $\alpha$ as the maximum value of $\alpha$, satisfying $\Delta \hat{\chi}^{2}=\hat{\chi}^{2}=2.3$.

\subsection{Test of SPL hypothesis}

\begin{table*}[!htb]
\centering
\scriptsize
\caption{The numbers of GRPs divided according to
the detection/non-detection in the individual bands.
The samples are classified with the consistency with SPL spectra
as well.}
\label{tcatMP}
\begin{tabular}{rrrrrrr}
\hline
 & & \multicolumn{2}{c}{SPL Hypothesis (MPGRP)}  & & \multicolumn{2}{c}{SPL Hypothesis (IPGRP)}\\
Detection Y/N & Group$^{a}$ & Consistent  & Inconsistent && Consistent  & Inconsistent\\
(P, LL, LH, S band)& &($\hat{\chi}_{\mbox{min}}^{2}<5.99$)    & ($\hat{\chi}_{\mbox{min}}^{2}>5.99$) && ($\hat{\chi}_{\mbox{min}}^{2}<5.99$)    & ($\hat{\chi}_{\mbox{min}}^{2}>5.99$)\\
\hline
\hline
(Y, Y, Y, Y)                     & I & 86 & 182 && 27 & 19\\
(Y, Y, Y, N)                     & I & 0  & 3 && 0 & 0\\
(Y, N, N, Y)                     & I & 0  & 1 && 0 & 0\\
(Y, Y, N, Y)                     & I & 0  & 0 && 0 & 0\\
(Y, N, Y, Y)                     & I & 5  & 11 && 1& 1\\
(Y, Y, N, N)                     & I & 0  & 0 && 0 & 0 \\
(Y, N, Y, N)                     & I & 0  & 0 && 0 & 0 \\
(N, Y, Y, Y)                     & II & 1230 & 619 ($\alpha>2$ : 9) && 88 & 34\\
(N, Y, N, Y)                     &II & 13 & 7 && 0 & 0\\
(N, N, Y, Y)                     &II & 342 & 60 ($\alpha>2$ : 4) && 15 & 1\\
(N, Y, Y, N)                     &\nodata & 0 & 24 && 0 & 1\\
(N, N, N, Y)                     & \nodata& 103 & 22 ($\alpha>2$ : 22) && 6 & 0\\
(N, Y, N, N)                     & \nodata& 0 & 2 && 0 & 2\\
(N, N, Y, N)                     & \nodata& 1 & 11 && 0 & 5\\
(Y, N, N, N)                     & III & 471 & 1 ($\alpha<-5$ : 1) && 52 & 1 ($\alpha<-5$ : 1)\\
\cline{2-7} \\
Total                                                  &  & 2251 & 943   && 189 & 64\\
                                                          &  &  &       ($\alpha>2$ : 35, $\alpha<-5$ : 1) && &($\alpha>2$ : 0, $\alpha<-5$ : 1)\\
\hline
\end{tabular}
\\ {\scriptsize a: We categorize the GRPs into three groups, from I to III. See Section \ref{cate}.}
\end{table*}

\begin{figure}[!htb]
\epsscale{0.8}
\plotone{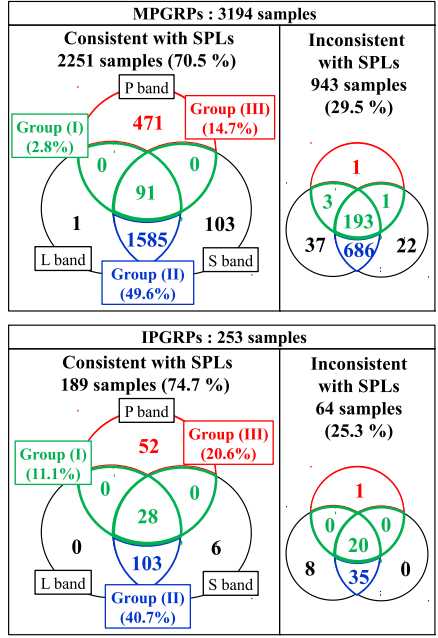}
\figcaption{Categorization of our spectral samples.
The numbers in the regions, where circles overlap each other mean the number of GRPs 
detected at corresponding multiple bands.
The GRPs detected at LL band and/or LH band are merged into the category of
the ``L band'' detection.
See also Table \ref{tcatMP}. \label{category}}
\end{figure}

\begin{figure}[!htb]
\epsscale{1}
\plotone{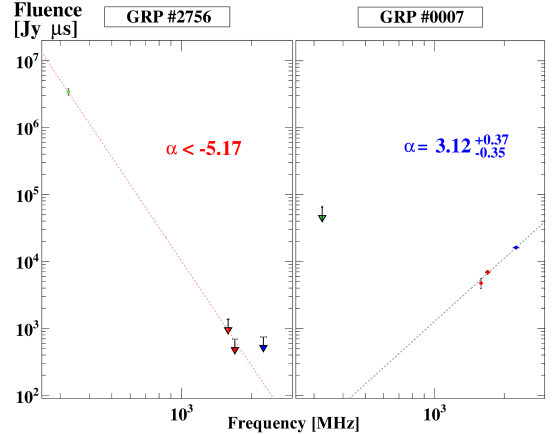}
\figcaption{Examples of the spectra of the "ultra-soft" (left) and the "ultra-hard" (right) GRPs.
For the ultra-soft GRP, the SPL function
with the 68\% confidence upper-limit of $\alpha$ is plotted
with the dotted line.
Also for the ultra-hard GRP,
the best-fit SPL function is plotted.
 \label{super}}
\end{figure}

\begin{figure}[!htb]
\epsscale{1.15}
\plottwo{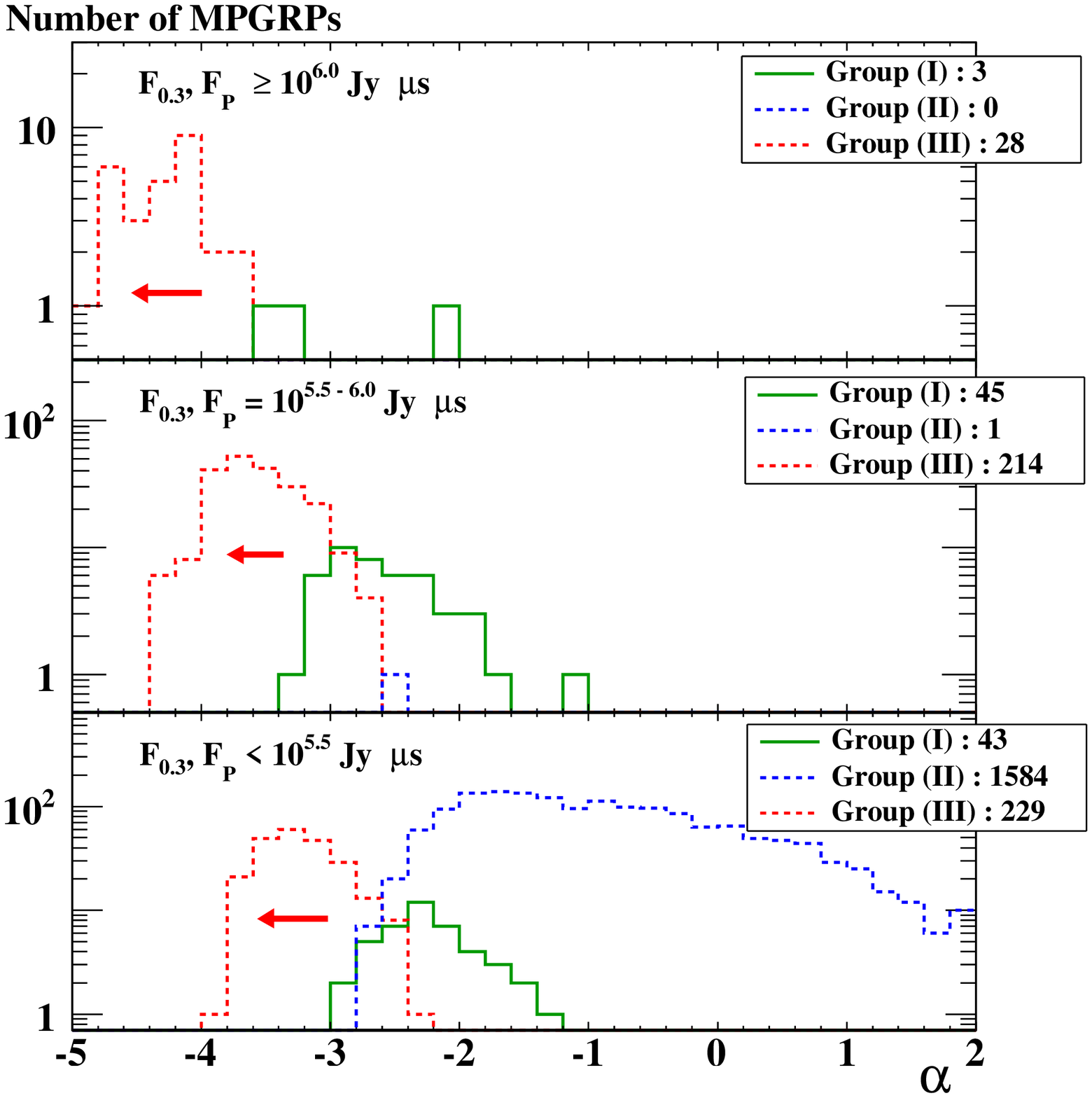}{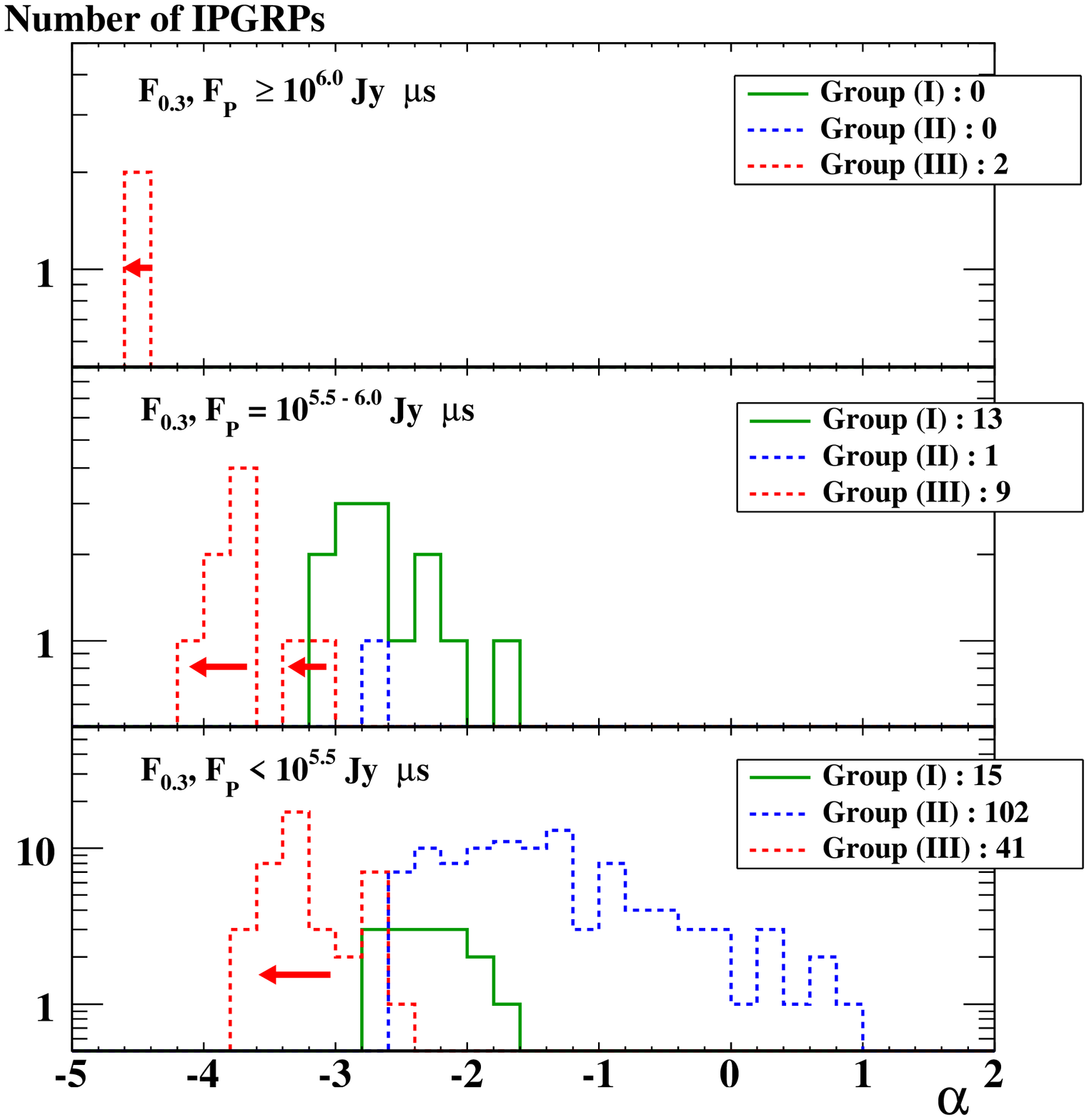}
\figcaption{Histograms of the spectral index $\alpha$ for the MPGRPs (left) and IPGRPs (right). 
Each panel is divided according to the fluence at P band:
$(F_{0.3},F_{\rm P}) \geq 10^{6} \mathrm{Jy}~\mu\rm{s}$ (top),
$10^{5.5} \mathrm{Jy}~\mu\rm{s} \leq (F_{0.3},F_{\rm P})< 10^{6} \mathrm{Jy}~\mu\rm{s}$ (middle),
and $(F_{0.3},F_{\rm P}) < 10^{5.5} \mathrm{Jy}~\mu\rm{s}$ (bottom).
The errors in $\alpha$ are neglected, and
the histogram of Group (III)
represents the upper-limit values;
i.e. the histograms with dashed lines (Group (II) and (III)) implies that
the values $\alpha$ are not determined firmly.
\label{alphahist-fludiv}}
\end{figure}

The numbers of the GRPs divided according to
the consistency of their spectra with SPLs, are summarized in Table \ref{tcatMP} and Figure \ref{category}.
In this analysis, GRP samples with an extreme spectral index,
$\alpha>2$ or $<-5$, are excluded from the SPL samples even if the $\hat{\chi}^{2}$-value is small enough.
Such examples are shown in Figure \ref{super}.
We find that the spectra of 2251 of 3194 MPGRPs (70.5\%) and 189 of 253 
IPGRPs (74.7\%) are consistent with SPLs at a significance level of 5\%.
We show the histograms of the spectral indices sorted by $F_{\rm P}$ or $F_{0.3}$
in Figure \ref{alphahist-fludiv}. Their mean spectral indices
of them are also shown in Table \ref{t3}. 

\begin{table}[!htb]
\caption{Mean spectral index $\alpha$
sorted by $F_{\rm P}$ or $F_{0.3}$.\label{t3}}
\begin{tabular}{ccrrr}
\hline
$F_{\rm P}$ or $F_{0.3}$ & Group & \multicolumn{3}{c}{$\alpha$}\\
\cline{3-5}
 &  & MPGRP & & IPGRP\\
 \hline \hline
                                                           & I       & $-3.03\pm0.61$   && \nodata\\
$\geq 10^{6.0}$ Jy~$\mu$s                                   & II      & \nodata               &&\nodata \\
                                                           & III      & $<-3.66$             &&$<-4.49$\\
                                                           & I        & $-2.58\pm0.44$  &&$-2.61\pm0.40$ \\
$10^{5.5} - 10^{6.0}$ Jy~$\mu$s                              &  II     & $-2.59$                &&$-2.61$\\
                                                           &  III    &$<-2.68$               &&$<-3.06$   \\
                                                           &  I       & $-2.23\pm0.37$  && $-2.29\pm0.31$\\
$< 10^{5.5}$ Jy~$\mu$s                                       &  II     & $-0.91\pm0.98$  &&$-1.38\pm0.82$\\
                                                            &  III   & $<-2.34$             &&$<-2.59$\\
\hline
                                                            &  I      & $-2.44\pm0.47$ &&$-2.44\pm0.38$\\
ALL                                                     &  II     & $-0.91\pm0.98$ &&$-1.39\pm0.82$ \\
                                                            &  III   & $<-2.34$             && $<-2.59$   \\
\hline
\end{tabular}
\end{table}

Note that our result does not necessarily imply that remaining
$\sim 30$\% of the GRP spectra totally deviate from an SPL.
As we mentioned before, the goodness of fit depends on the method of the error estimate (see Appendix \ref{binning1}).
Although we cannot conclude that the majority of GRP spectra are definitely SPLs,
there is no strong evidence that rejects SPLs as the typical spectral shape.
In Sections \ref{relation} and \ref{corLS},
we discuss the GRP spectra consistent with SPLs,
and then discuss the inconsistent ones in Section \ref{inconsis}.

\subsection{The P-band Fluence and Hardness}
\label{relation}

Similarly to the result in \cite{SB99}, we do not find
a significant correlation between the fluences at P and L bands.
This suggests that the spectral index distributes widely.
We investigate the distributions of the index $\alpha$
as a function of the fluence below.

\begin{figure}[!htb]
\epsscale{1.0}
\plotone{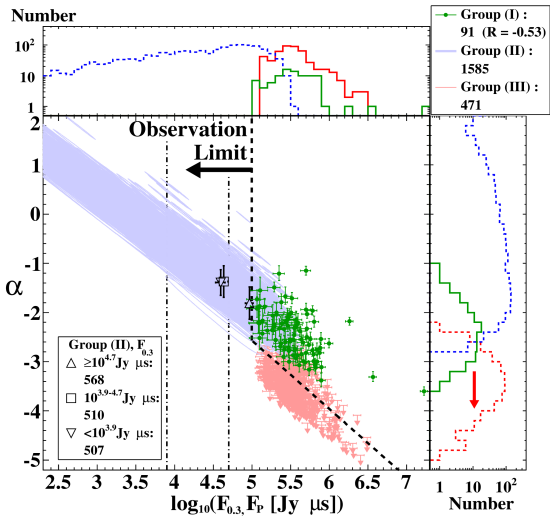}
\figcaption{
Scatter plot of the observed/extrapolated P-band fluences
and the spectral indices $\alpha$
for the MPGRPs whose spectra are consistent with SPLs.
The data for Groups (I), (II), and (III) are shown with green, blue,
and red symbols, respectively.
The upper-left and the lower-right panels are the number distributions
for the parameters of their axes,
where we put samples into a bin neglecting the errors.
Note that the $\alpha$-histogram for Group (III) 
consists of the upper-limit values.
In the upper-right panel, $R$ is the Spearman rank correlation coefficient for Group (I).
The dashed lines represent the observation limit to judge a GRP as one of Group (I).
By the vertical dash-dotted lines in the central panel,
we further divide the samples in Group (II) into three sub-groups.
The average values of $F_{\rm P}$ and $\alpha$
($\bar{F}_{\rm{stk}}$ and $\bar{\alpha}_{\rm{est}}$)
obtained with the stacking analysis
for each sub-group
are plotted with the open triangles and the square
(see Section \ref{relation} and Table \ref{tgroup2}).
\label{flualpha}}
\end{figure}

\begin{figure}[!htb]
\epsscale{1}
\plotone{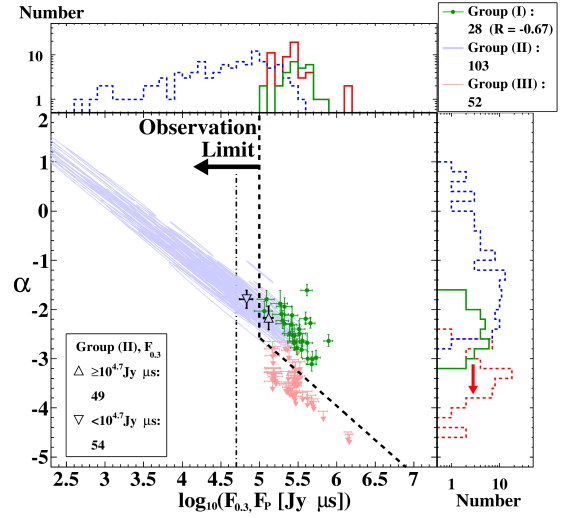}
\figcaption{Same plot as Figure \ref{flualpha}, but for the IPGRPs. \label{flualpha-IP}}
\end{figure}

Since the most of the radiation energy is released at P band or lower frequencies
for an SPL GRP with $\alpha<-1$, we focus on the P-band fluence here.
For the samples consistent with SPLs,
we show the scatter plots in the parameter space of SPL in the central panel of
Figures \ref{flualpha} and \ref{flualpha-IP} for the MPGRPs and IPGRPs, respectively.
In those figures, we plot the observed fluences
$F_{\rm P}$ for Groups (I) and (III), while the values $F_{0.3}$
estimated from Equation (\ref{fit0.3}) are plotted for Group (II).
With the 68\% confidence interval, the errors in the figures are plotted
in the standard manner for Group (I),
but the confidence region ellipses \citep[][]{Pr92}
are directly drawn for Group (II).
This is because the allowed regions for Group (II) are inclined and elongated,
as shown in the middle panel of Figure \ref{Chimap},
reflecting the fact that the value of $F_{0.3}$ estimated from extrapolation largely
depends on $\alpha$.
For Group (III), we plot the upper limits on $\alpha$
with the observed $F_{\rm P}$.

At a glance, the scatter plots seem to show a correlation between the hardness and fluence.
Figure \ref{alphahist-fludiv} also seems to show the tendency
of softening with increasing fluence (brighter--softer).
However, we should take into account the observational bias.
In the scatter plots, we also draw the rough ``observation limit'' lines
(the limit depends on the pulse duration technically)
indicating the regions where the fluence at P or LH band
\footnote{For an SPL-GRP with $\alpha \lesssim -3$,
the detectability above P band is the highest at LH band in our observational configuration.}
is below the detection limit.
This limit corresponds to the necessary condition
to classify an SPL-GRP into Group (I).
In the scatter plots, 
dimmer and softer GRPs are not found in the lower-left region.
Such GRPs are also expected to be dim at L/S bands so that
the deficit of GRPs in that region can be due to the observation bias.
On the other hand, the brighter and harder (the upper-right region) GRPs 
are expected to be easily detected at the L/S band as long as the spectrum extends.
We extract only the samples in Group (I) (green points),
whose parameters are determined most accurately, and
calculate the Spearman rank correlation coefficient, $R$ \citep[e.g.][]{Pr92}.
The values of $R$ are $-0.53$ for the MPGRPs and $-0.67$ for the IPGRPs.
From the values of $R$,
the probabilities that those two variables are unrelated (hereafter $p$-value) are $7\times10^{-8}$ for MPGRPs, and $1\times10^{-4}$ for IPGRPs.
However, the observational limit, denoted with the inclined lines, may affect those values.

It is also seen from those scatter plots that, whereas the samples in Group (II) imply an existence of a significant number of
hard GRPs ($\alpha \gtrsim -1$), there are not many hard GRPs in Group (I)
(see also Figure \ref{alphahist-fludiv}).
The Group (II) GRPs are not detected at P band consistently with the estimated hard spectral index.
This also seems to support the apparent correlation, brighter--softer or dimmer--harder.
Since the fluences at P band for Group (II) are extrapolated values,
it is worth confirming whether the average flux after stacking over the corresponding GRP intervals
exceeds the normal pulse flux.
We collect the light curves at the P band
in the rotational periods when Group (II) GRPs occurred.
According to the extrapolated values of $F_{0.3}$,
the MPGRPs are divided into three sub-groups  with
two boundary values at $F_{0.3}=10^{3.9}$ and $10^{4.7}$Jy~$\mu$s.
Similarly, the IPGRPs are divided into two sub-groups with a boundary value at $F_{0.3} =10^{4.7}$Jy~$\mu$s.
These boundary values are chosen so as to make the sub-groups of similar size
(Table \ref{tgroup2}).

\begin{figure}[!htb]
\epsscale{1.15}
\plottwo{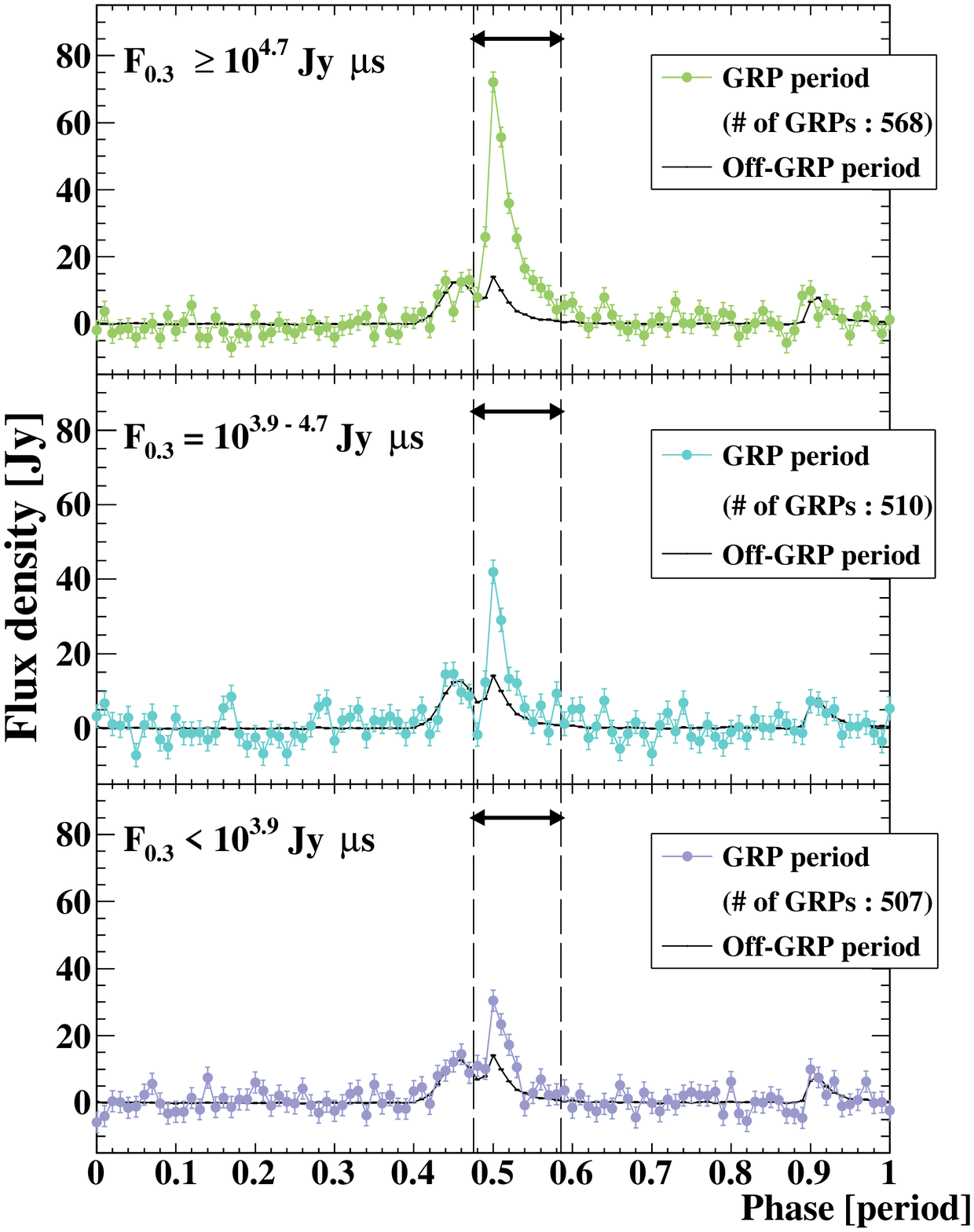}{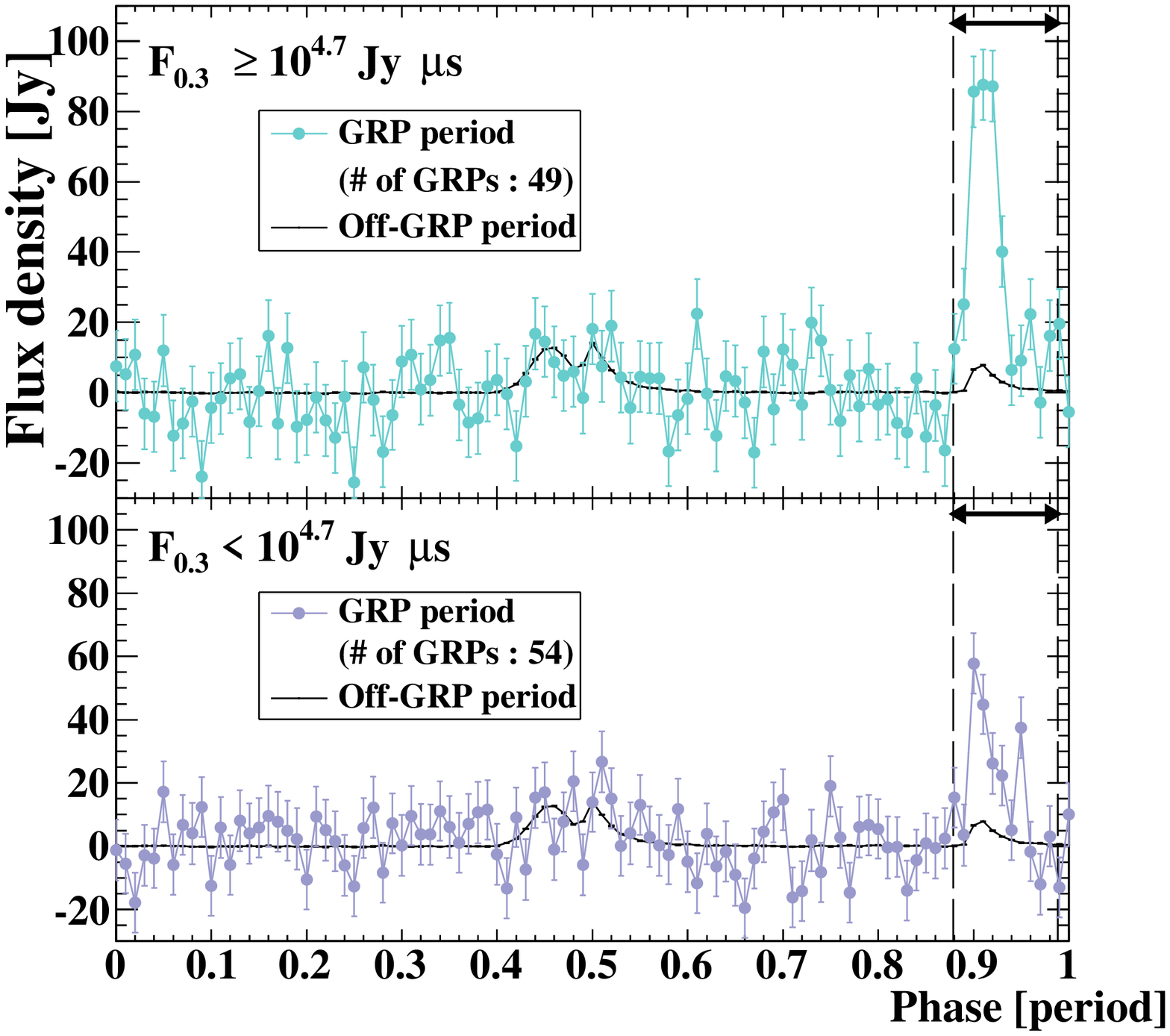}
\figcaption{Average pulse profiles at the P band, when Group (II) MPGRPs (left) and IPGRPs (right) occurred.
The samples are divided into sub-groups according to the fluence $F_{0.3}$.
We also show the average profile for the off-GRP periods.
The $1\sigma$ statistical errors
are shown with the error bars.
The phase intervals to estimate the fluence
are shown by the arrows. \label{g2stackMP}}
\end{figure}

The resultant average pulse profiles
are shown with the colored lines in Figure \ref{g2stackMP}.
Compared to the average pulse profile of the off-GRP periods,
it is clearly seen that the pulse at the P band is on average
enhanced at the GRP period defined with L/S bands.
The average fluences $\bar{F}_{\rm{stk}}$ and $\bar{F}_{\rm{off}}$
for on- and off-GRP periods, respectively,
are evaluated for phase intervals
shown with the arrows in Figure \ref{g2stackMP}.
The width of the intervals is 0.11 period, 
which is the sum of the width for the GRP selection
(0.05 period) and the characteristic duration of the GRPs
at the P band ($\sim 2$ms, 0.06 period).
The calculated values of $\bar{F}_{\rm{stk}}$ and $\bar{F}_{\rm{off}}$
are summarized in Table \ref{tgroup2}.
From $\bar{F}_{\rm{stk}}$ and the fluence $F_{\rm{S}}$ of each GRP at S band,
we deduce average spectral indices $\bar{\alpha}_{\rm{est}}$ as
\begin{equation}
\bar{\alpha}_{\rm{est}}\equiv\frac{1}{N_{\rm{GRP}}}\sum_{i}^{N_{\rm{GRP}}}\frac{\log{(\bar{F}_{\rm{stk}}/F_{\rm{S},i}})}
{\log{(325.1\rm{MHz}/2250\rm{MHz})}}.
\end{equation}
The values of $\bar{\alpha}_{\rm{est}}$ 
are also tabulated in Table \ref{tgroup2},
and $(\bar{F}_{\rm{stk}},\bar{\alpha}_{\rm{est}})$ are plotted
in Figures \ref{flualpha} and \ref{flualpha-IP} as the open black symbols.
Those average points are consistently on the clusters of the blue ellipses,
which are extrapolated from fluences at L and S bands,
and seem to reinforce the SPL hypothesis and the tendency of the softening with increasing $F_{\rm P}$,
considering the green data points of Group (I).
In addition, the obtained values of $\bar{\alpha}_{\rm{est}}$ are significantly
larger than the average values for the samples with brighter $F_{\rm P}$,
Groups (I) and (III) (see Figure \ref{alphahist-fludiv} and Table \ref{t3}), or the index of the normal pulse ($\sim -3$).

\begin{table*}[!htb]
\begin{center}
\caption{The mean fluences $\bar{F}_{\rm{stk}}$, $\bar{F}_{\rm{off}}$ and
spectral indices $\bar{\alpha}_{\rm{est}}$.}
\label{tgroup2}
\begin{tabular}{rcccccc}
\hline
&\multicolumn{3}{c}{MPGRP} & {} & \multicolumn{2}{c}{IPGRP}\\
\cline{2-4} \cline{6-7}
$\log_{10}F_{0.3}[\rm{Jy~\mu s}]$ & $\geq 4.7$ & $3.9-4.7$ & $<3.9$ & &$\geq 4.7$ & $<4.7$ \\
\hline\hline
$N_{\rm{GRP}}$\tablenotemark{a} & 568 & 510 & 507 & & 49 & 54 \\
$\log_{10}\bar{F}_{\rm{stk}}[\rm{Jy~\mu\rm{s}}]$ & $4.97^{+0.05}_{-0.05}$ & $4.64^{+0.05}_{-0.07}$ & $4.60^{+0.05}_{-0.07}$ & & $5.12^{+0.05}_{-0.07}$ & $4.83^{+0.08}_{-0.09}$ \\
$\log_{10}\bar{F}_{\rm{off}}[\rm{Jy~\mu\rm{s}}]$ & \multicolumn{3}{c}{$4.28^{+0.05}_{-0.05}$} & & \multicolumn{2}{c}{$3.98^{+0.05}_{-0.05}$} \\
\hline
$\bar{\alpha}_{\rm{est}}$ & $-1.81\pm0.33$ & $-1.37\pm0.32$ & $-1.39\pm0.26$ & & $-2.17\pm0.24$ & $-1.79\pm0.18$ \\
\hline
\multicolumn{2}{l}{a. Number of GRP samples.}&&&&&\\
\end{tabular}
\end{center}
\end{table*}

For the brighter sub-groups of $F_{0.3}\geq 10^{4.7}\rm{Jy}~\mu\rm{s}$,
the obtained values of $\bar{F}_{\rm{stk}}$ are consistent with the
prospective fluence $F_{0.3}$.
However, for the dimmer sub-groups, $F_{0.3}<10^{3.9}\rm{Jy}~\mu\rm{s}$ for MPGRPs
and $F_{0.3}<10^{4.7}\rm{Jy}~\mu\rm{s}$ for IPGRPs,
the values of $\bar{F}_{\rm{stk}}$ become higher than the corresponding range of $F_{0.3}$.
If the SPL hypothesis is correct even for such dimmer GRPs,
the above contradiction may be due to the contribution of the normal pulse.
Since we have assumed that the normal pulse is absent at the GRP periods,
we have not subtracted the contribution of the normal pulse.
For such low-fluence GRPs, the method to distinguish
GRPs from the normal pulses is not established so far.
Another interpretation for the discrepancy between $\bar{F}_{\rm{stk}}$ and $F_{0.3}$
is that non-SPL GRPs are contaminated in Group (II) samples.

\subsection{The L/S-band Fluence and Hardness}
\label{corLS}

In the previous subsection, we have found signatures of a possible correlation
between the fluence at the P band and hardness.
However, the observational bias obscures the conclusion for the correlation.
To probe clues about the correlation from another angle,
we discuss the correlation between the L/S band fluence and hardness in this subsection.

\begin{figure}[!htb]
\epsscale{1.15}
\plottwo{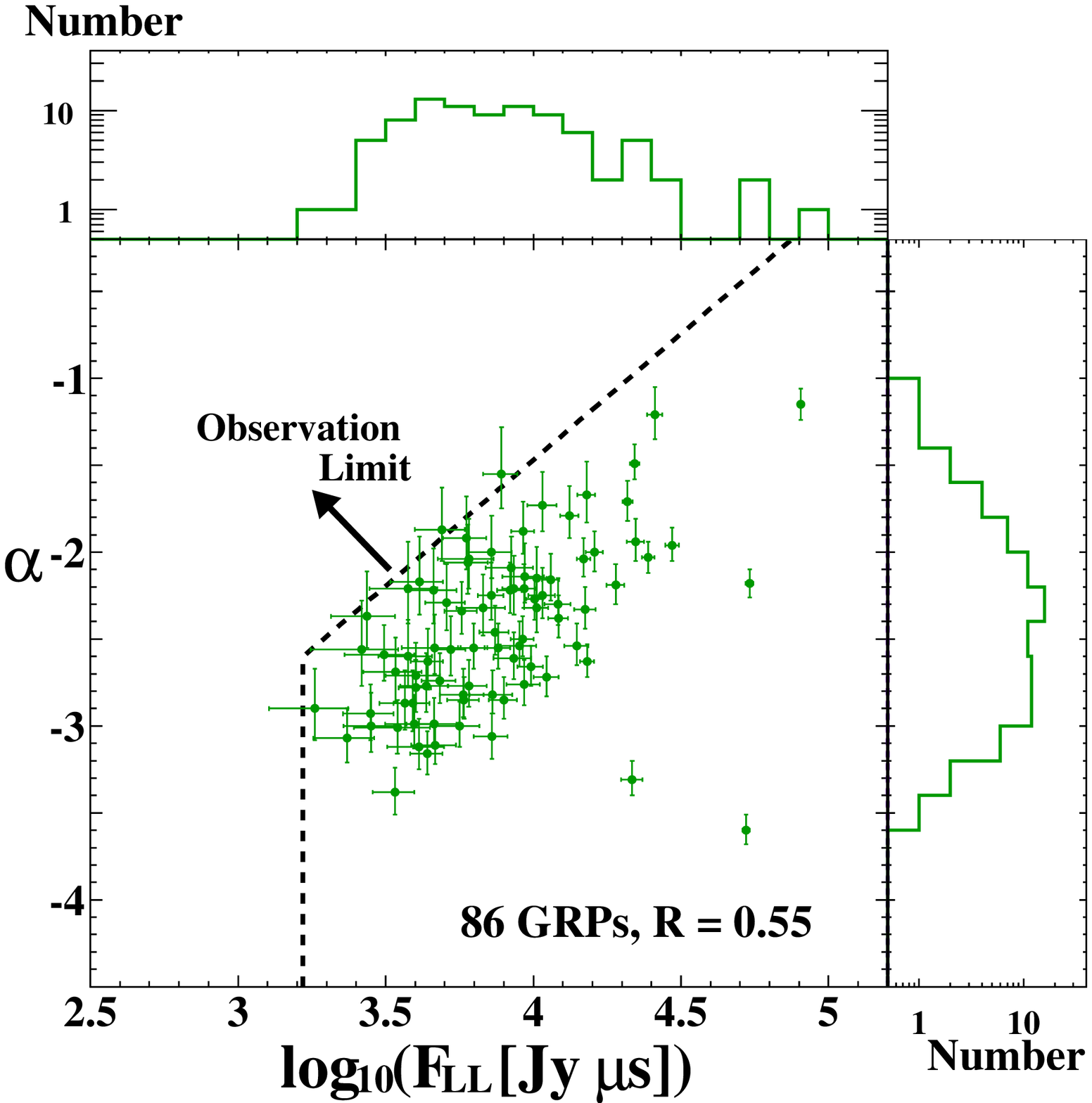}
{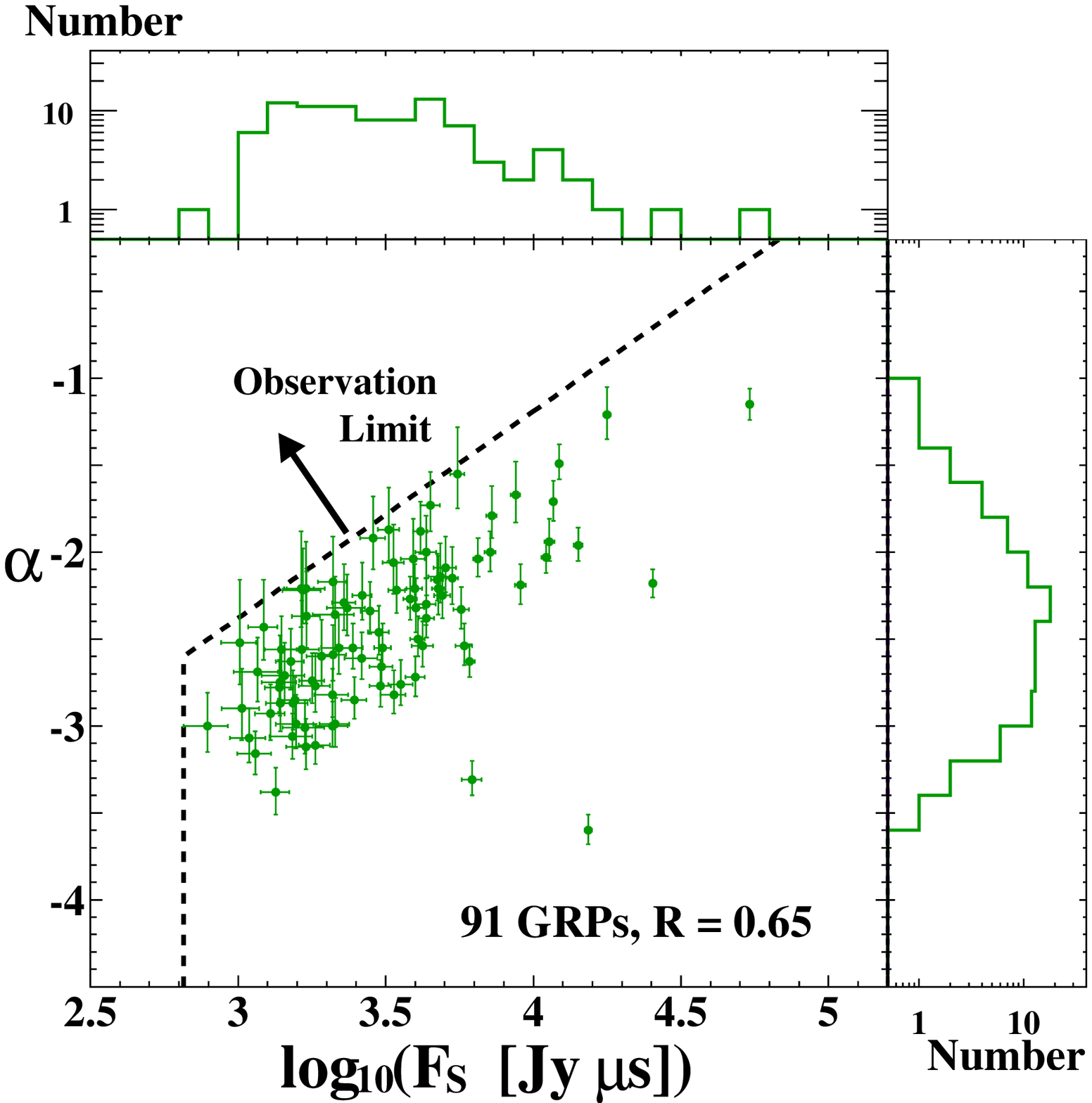}
\figcaption{Scatter plots of the observed fluences
at LL (left) and S (right) bands and the spectral indices $\alpha$
for the MPGRPs belonging to Group (I).
The distributions of the fluences and indices are projected on 
the upper and right sides of each panel.
The dashed lines represent the observation limit to be detected at both P and corresponding bands.
The Spearman rank correlation coefficients $R$ are also shown. \label{flualpha-LS}}
\end{figure}

\begin{figure}[!htb]
\epsscale{1.15}
\plottwo{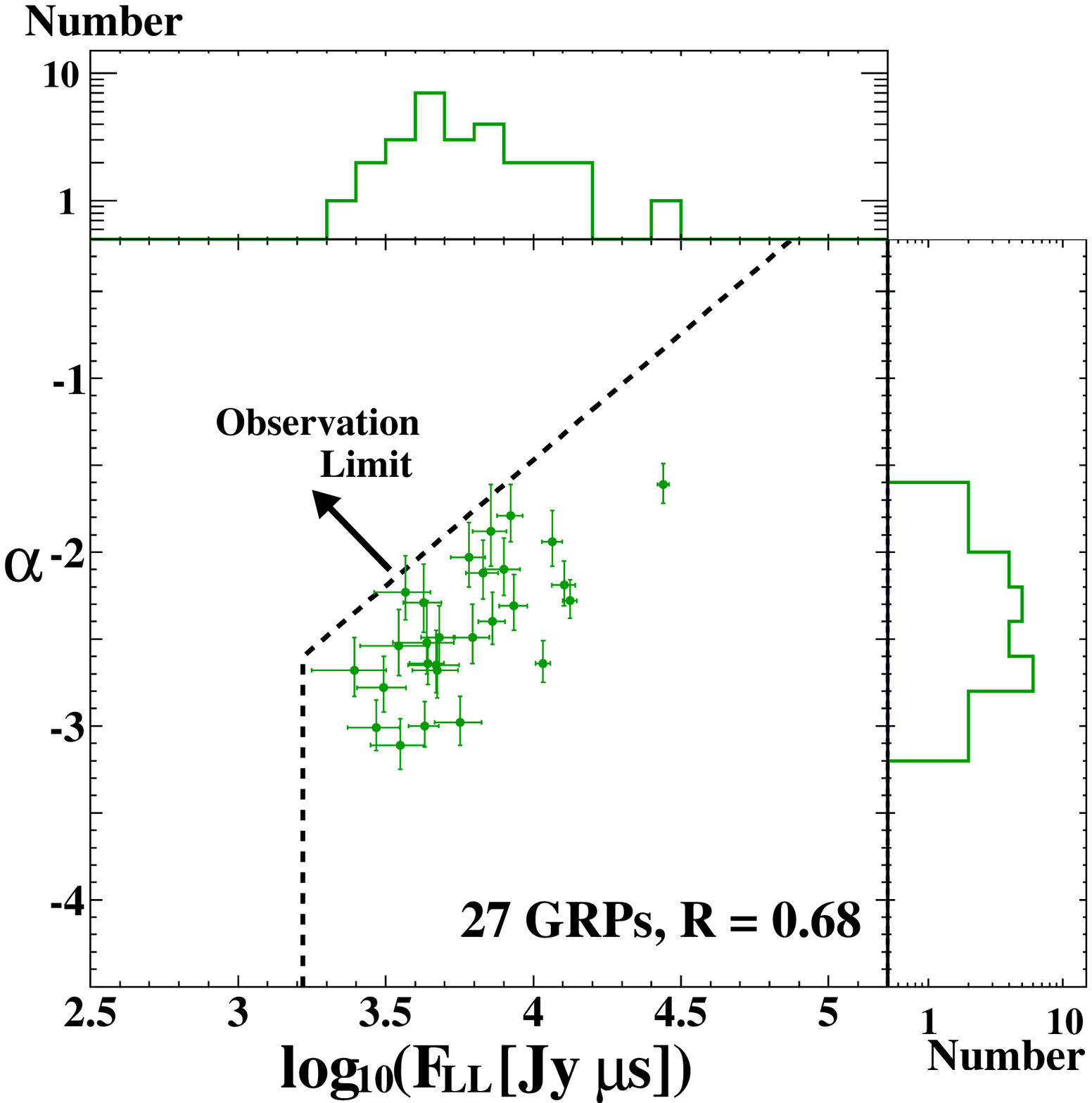}
{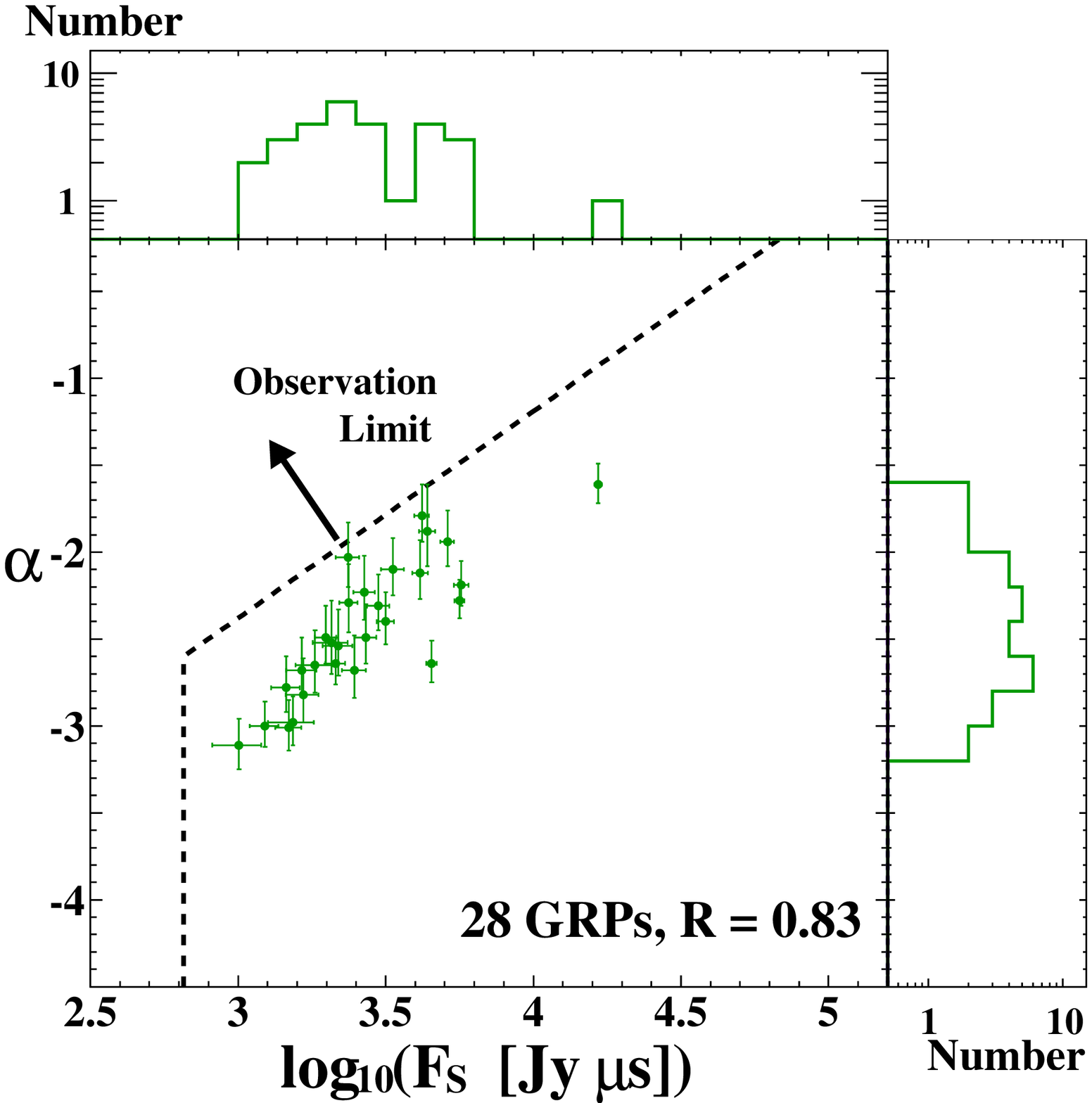}
\figcaption{Same as Figure \ref{flualpha-LS} but for IPGRPs.
\label{flualpha-LS_IP}}
\end{figure}

We choose the GRPs, whose spectra are consistent with SPLs
in Group (I)
(see Tables \ref{tcatMP} and Section \ref{cate}).
The uncertainty in the spectral index are relatively low
owing to the wide frequency separation between P and L/S bands
(see Section \ref{cate}).
We show scatter plots of the fluences at LL or S band and the indices $\alpha$
in Figures \ref{flualpha-LS} and \ref{flualpha-LS_IP} for the MPGRPs and IPGRPs, respectively.
Since the detection limit becomes deepest at the LH band,
the fluence limit at the LL band is the fluence at the LL band
for an SPL GRP, whose fluences at LH and P bands are on the detection limit.
The apparent correlations between the fluence and hardness
are seen clearly.
From the Spearman rank correlation coefficients,
the $p$-values for the correlation absence described in the previous subsection are
$4\times10^{-8}$ and $3\times10^{-12}$ 
for MPGRPs at LL and S bands, respectively.
While we have omitted the results for LH band,
the results are similar (91 samples, $R=0.57$ for MPGRPs
and 28 samples, $R=0.79$ for IPGRPs).
The deficit of samples in the upper-left region
can be due to the selection bias.
However, the deficit in the lower-right region cannot be explained
by the selection bias.

Contrary to the case of the P band, a positive correlation
between the fluence and spectral hardness is found in the cases of L/S bands.
If the negative correlation in the P-band case is real,
our results suggest that
the fluences at P band and L/S band have an anti-correlation.
This is intuitively understood from Figure \ref{sp4}.
Since the most of the GRP energy is released at P band or lower frequencies,
the anti-correlation means that more energetic GRPs tend to show
softer spectra.


We carry out Monte Carlo simulations to reproduce the scatter distributions
without an intrinsic correlation between the L-band fluence and hardness
(see Appendix \ref{montecarlo}).
When we assume that the typical spectral index is close to that of the normal pulse of $-3$,
the Spearman coefficient in the $F_{\rm LL}$--$\alpha$ diagram cannot be reproduced.
However, if we assume the typical index of $-1$, harder than the normal pulse's one,
the apparent correlations in both the $F_{\rm LL}$--$\alpha$ and $F_{\rm P}$--$\alpha$
are reproduced. This is consistent with the hard spectrum implied from the stacking analysis.
Given $F_{\rm LL}$, a smaller $\alpha$ implies a brighter P-band fluence,
where most of energy is released.
Namely, the uncorrelation between $F_{\rm LL}$ and $\alpha$ is equivalent
to an anti-correlation between the total energy release and hardness.
Therefore,
the apparent correlations in Figures \ref{flualpha}, \ref{flualpha-IP},
\ref{flualpha-LS}, and \ref{flualpha-LS_IP} imply that a more energetic GRP tends
to have a softer spectrum.
In our samples, the P-band GRPs, including Group (III) GRPs,
whose spectrum should be very soft, are energetic samples compared to
Group (II) GRPs.
The different average $\alpha$ for P band GRPs and Group (II) GRPs
also indicates the anti-correlation between the energy release and hardness.


When we add the GRPs in Group (II) for the correlation analysis,
a significant correlation is not found.
This may be partially due to the large errors in the index of Group (II) GRPs.
Alternatively, another type of GRPs, whose spectra
are not SPLs may be contaminated in Group (II) samples.

\subsection{Inconsistent Spectra with SPLs} 
\label{inconsis}

The fractions of the GRPs judged to be inconsistent with SPLs
at a significance level of 5\% are 28.4\% and 24.9\% for the MPGRPs and the IPGRPs, respectively.
When the significance level is set to 0.1\% ($\hat{\chi}^2<13.82$),
the fractions are 14.2\% and 14.6\% for the MPGRPs and the IPGRPs, respectively.
Examples of the spectra apparently inconsistent with SPLs
are shown in Figure \ref{chisqhigh}.
Some spectra show softening or hardening at higher frequencies.
Such hard-to-soft (left), soft-to-hard (center), and other (right) examples
are displayed in the figure. 
\citet[][]{Po08,Po09} claimed that GRP spectra tend to 
show soft-to-hard behavior at 0.6--2.2 GHz.
In order to examine this tendency,
we choose 182 MPGRPs and 19 IPGRPs that are detected
at all the bands from P to S, and are inconsistent with SPL spectra
at a significance level of 5\%.
In those MPGRP samples, the fractions of hard-to-soft, soft-to-hard, and other
spectra are 48 (26\%), 76 (42\%), and 58 (32\%), respectively.
For the 19 IPGRPs, the numbers are 3 (16\%), 4 (21\%), and 12 (63\%), respectively.
We do not find a clear tendency in the spectral behavior.

We find 1 (0.03\%) ultra-soft ($\alpha<-5$) and 35 (1.1\%) ultra-hard ($\alpha>2$) MPGRPs,
and we also find 1 (0.4\%) ultra-soft IPGRPs (see Figure \ref{super}).
In our GRP samples, no ultra-hard IPGRP is found.

\section{GRPs Detected at the C/X Band}
\label{5band}
In this section, we discuss the GRPs detected at C and/or X band(s).
Note that the fluences at C and X bands are expected to be affected
by intensity modulation due to the interstellar scintillation.

We find 5 MPGRPs and 19 IPGRPs exceeding the selection thresholds
(see Section \ref{GRPsel} and Table \ref{t1}) at C and/or X band(s).
All of these five MPGRPs were simultaneously detected
at some of the frequency bands of P, L, or S band.
In contrast, all the 19 IPGRPs were detected only at X band.
As for the IPGRPs in Groups (I) and (II)
discussed in Section \ref{3band},
the spectra may have a cut-off below the X band frequency.

\begin{figure}[!htb]
\epsscale{1}
\plotone{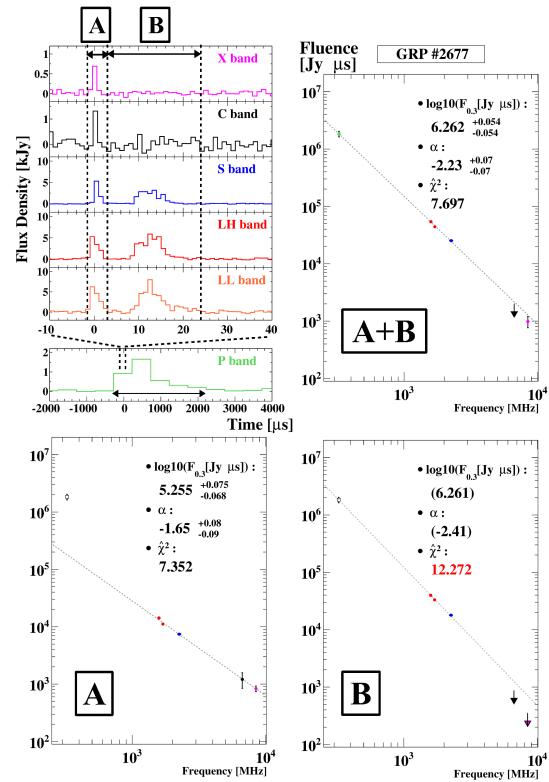}
\figcaption{Spectra of the GRP (upper-right), whose dynamic spectra and pulse profiles
are shown in Figures \ref{dynsp1} and \ref{sp1}.
The spectra for time intervals A and B defined in the upper-left light curve
are shown in the bottom panels.
The data point at P band is excluded for the fits for the time intervals A and B, respectively.
 \label{golden}}
\end{figure}

In Figure \ref{golden}, we show the spectrum of the MPGRP \#2677,
whose pulse profiles and dynamic spectra are
shown in Figures \ref{dynsp1} and \ref{sp1}, respectively.
The spectrum is consistent with a SPL at P--X bands at a significance level of 5\%.
This is only the case consistent with a SPL in the five MPGRPs.

As seen in Figure \ref{golden}, the light curve shows two distinct components,
so that we attempt to obtain the temporally divided spectra here.
When we divide the light curve into two intervals A and B defined in Figure \ref{golden},
it turns out that the spectrum for the interval B needs a high-frequency cut-off.
In Figure \ref{sp1}, many finer (sub-$\mu$s) structures \citep[e.g.][]{Ha03}
are found in the light curve.
The spectrum may consist of further more components as well.

\begin{figure}[!htb]
\epsscale{1}
\plotone{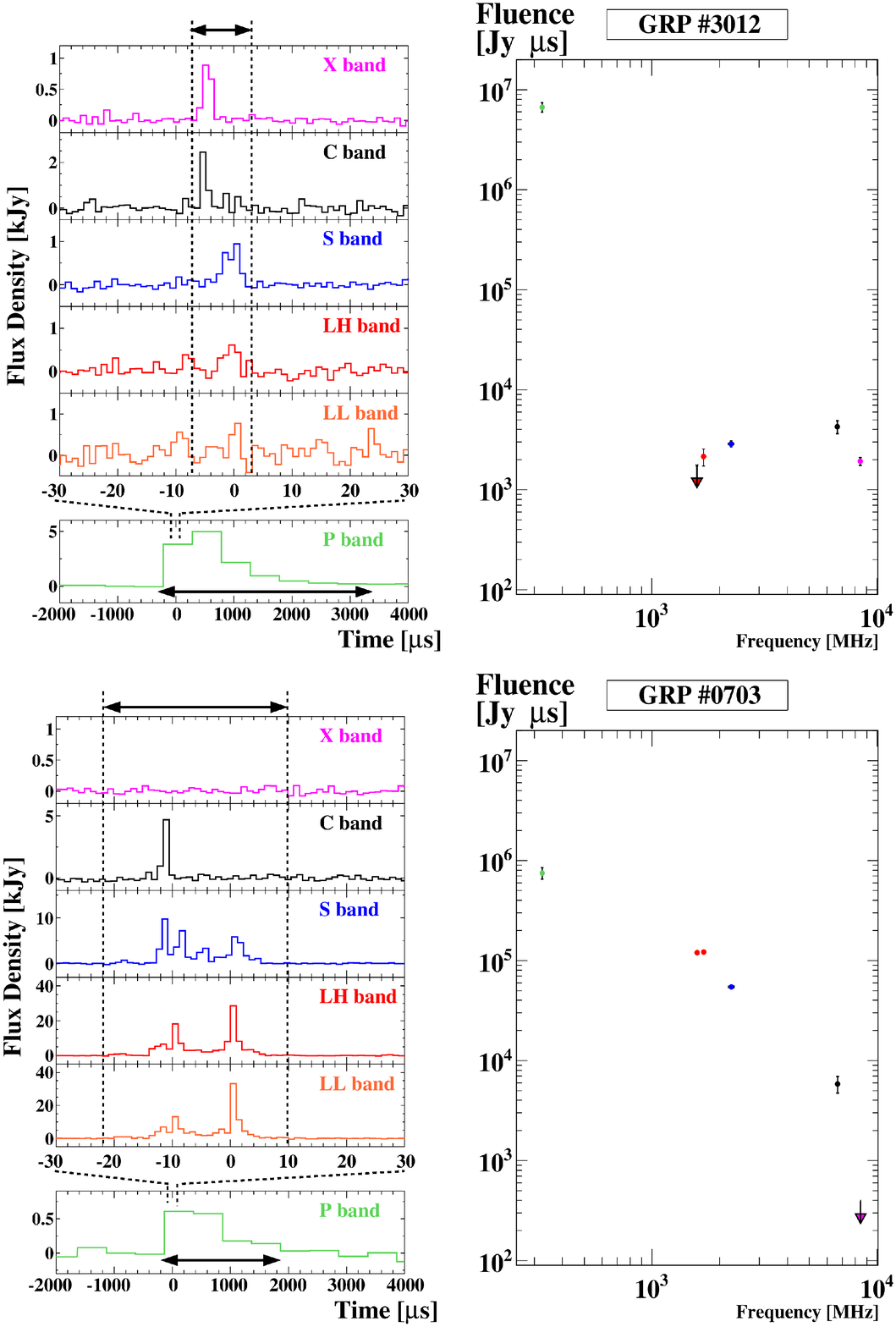}
\figcaption{Examples of broadband GRP spectra (right) inconsistent with SPLs.
The pulse profiles (left) are also shown with
the time intervals for the integral.
\label{bronze}}
\end{figure}

The spectra of the other four MPGRPs are inconsistent with SPLs
as shown in Figure \ref{bronze}.
The P-band fluence of GRP \#3012 is about three orders of magnitude brighter
than fluences at the higher frequency bands.
The spectrum seems to consist of two or more different components,
which may be supported by the slight misalignment of the peak times
($\sim4\ \mu$s) between the light curves in S and C/X bands. 
The small uncertainty of the TOA
between LL and X bands ($\sim1.5\ \mu$s) suggests that
the $4\ \mu$s misalignment is an intrinsic feature.
The pulse profile of GRP \#0703 appears
to have multiple components
 (especially at L and S bands),
similarly to the GRP \#2677 (Figure \ref{golden}).
The stringent upper-limit at the X band requires
a sharp high-frequency cut-off in the spectrum.
For GRP \#0703, 
the temporally divided spectra are also inconsistent with SPLs
and need low/high-frequency breaks or cut-offs.

\section{Discussion and Conclusion} \label{dis}

The broadband (0.3, 1.6, 2.2, 6.7, and 8.4 GHz) simultaneous observation
of the GRPs from the Crab pulsar was conducted
with four Japanese observatories, Iitate, Kashima, Usuda, and Takahagi,
on 2014 September 6-7th. 
We obtained 3194 MPGRPs and five of them were detected
at the multiple frequency bands including 6.7 and/or 8.4 GHz.
We also obtained 272 IPGRPs and 19 of them were detected
only at 8.4 GHz. We investigated the spectra of those GRPs.
Our results are as follows.
\begin{itemize}
\item In the frequency range from P to S bands (0.3--2.2 GHz),
70\% or more of the spectra in our GRPs are consistent with SPLs.
We do not find a strong evidence that the majority of GRPs
disagree with SPL spectra.

\item For the GRPs, whose spectra are consistent with SPLs,
the spectral index is widely distributed from approximately $-4$ to $-1$.

\item The GRP distributions in the fluences at P band and spectral index
show a possible negative correlation.

\item For the L/S band GRPs, which are not detected at P band,
we carry out the stacking analysis of the P-band light curve.
We find significant enhancement in the P-band light curves
compared to the light curves in the non-GRP periods.  
This indicates that those L/S-band GRPs
accompany dim P-band GRPs.
The average spectral indices for those samples,
approximately $-1$ or $-2$, are harder than 
the bright GRPs.
This supports the possible negative correlation between the P-band flux and
hardness.
Since the dim and hard GRPs (Group (II)) occupy about a half of our GRP samples,
it might be suggested that most of GRPs may have intrinsically harder spectra
than that suggested in the previous studies.

\item A positive correlation between the L/S band fluence and hardness
is found. Considering the negative correlation at P band,
our results suggest an anti-correlation between the total energy release
and hardness.
Our Monte Carlo simulations to reproduce the distributions
of the fluences and $\alpha$ show that
the typical spectral index is as hard as $-1$ again.

\item We find several ultra-soft ($\alpha<-5$) and ultra-hard ($\alpha>2$) GRPs.

\item A small fraction of spectra apparently deviate from SPLs.
We do not find any tendency in the spectral shape for such GRPs.

\item Several GRPs are detected in a very wide frequency range
of $0.3$--$8.4$ GHz. Most of them need multiple spectral components.

\end{itemize}
\subsection{Comparison with the previous studies}
\citet{Ka10} reported that the distribution of the spectral index
is extremely wide
(from about $-15$ to $+10$) for a narrow frequency range of 1.3-1.45 GHz.
In our observation, the fluences at the four channels in L band
significantly fluctuate (see Appendix \ref{binning1}). If we fit such data with an SPL,
a similar result to \citet{Ka10} may be obtained.
Such fluctuation might be intrinsic, or
due to the scintillation modulation.
It should be noticed that a large fraction of our broadband spectra
do not conflict with SPLs from P to S bands.
Our broadband analysis seems more convincing than the results
based on the narrow frequency range.
 
\citet{SB99} investigated the spectral index between 0.6 and 1.4 GHz,
and obtained their distribution ranging from $-4.9$ to $-2.2$.
Although the frequency range is slightly different,
the index for Group (I)
is distributed within a similar range
(see Figure \ref{alphahist-fludiv}). 
\citet[][]{Or15} investigated the spectral index between 0.2 and 1.4 GHz,
and obtained their distribution ranging from $-4.9$ to $-3.6$.
However, the fluences at 1.4 GHz in \citet[][]{Or15}
are about two orders of magnitude smaller than those in the previous studies
\citep[e.g.][]{SB99,Bh08}, which are roughly consistent with our values,
so that the results in \citet{Or15} should be reconsidered. 

\subsection{Emission Mechanism}
\label{theoGRP}
Our results suggest the anti-correlation
between the total energy release and hardness.
The emission mechanism for the broadband GRPs
may be different from the anomalous cyclotron resonance model
\citep{Ly07}
proposed for the characteristic property of the IPGRPs above 4GHz \citep{HE07},
for which we did not detect the signal of the counterpart at P--S bands.
In the induced scattering model \citep{Pe04}, the intensity
variation is attributed to the fluctuation
of the optical depth to the induced scattering.
\citet{Pe04} suggested that the intensity amplification for the Crab GRPs
is more efficient at the lower frequencies, which agrees with the correlation we found.
Alternatively, a larger energy release may enhance the efficiency
of the electron--positron pair production.
The resultant high-density pairs may screen out the electric field relatively soon,
which may reduce the fraction of higher energy electrons/positrons.
Considering the unstable behavior of the magnetosphere as demonstrated by
\citet{Tim13} or \citet{kis16},
the above simple interpretation can be an option to explain the correlation.
In any case, before jumping to a conclusion,
we need to examine various nonlinear processes
in the GRP activity such as
the detailed properties of the charge bunching due to two stream instability \citep{Che80}
and the photon softening due to the induced Compton scattering
\citep[e.g.][]{Tan15}.

\subsection{Comparing with FRBs}
Extremely bright GRPs from young pulsars are candidates of the emission process
for FRBs \citep[e.g.][]{CW15,Co15}.
The measured spectrum indices of FRBs are from $\sim-4$ to $\sim+1$
\citep{Lo07, Ke12, Ra15, Ke16}, which are roughly consistent with that of Crab GRPs.
One exception is the repeating FRB
121102, whose range of the spectrum index is extremely wide, from $\sim-10$ to $\sim+14$
\citep{Sp16}.
Although such hard and soft bursts may correspond to the ultra-hard and
ultra-soft GRPs in our samples (Figure \ref{super}), this wide range of the spectral
index may be due to the narrow frequency range (1.2--1.5GHz) in their observation.
In fact, if we obtain the spectral index distribution from
fluences in LL and LH bands, the range is from $\sim-15$ to $\sim+10$ \citep[see also][]{Ka10}. The wide frequency observations have the potential
to provide unique limits on the spectral index distribution of FRBs \citep[e.g.][]{Bu16}. The upcoming Canadian Hydrogen Intensity Mapping
Experiment \citep[CHIME,][]{Ba14} will obtain a large number of FRBs
($\sim 10$ per day) in a wide frequency range of 400--800 MHz.
Such observations \citep[see also][for FRB search with VLBI]{Tak16} can provide a great opportunity to probe the possible correlations
between the spectral index and the fluence similar to the Crab GRPs.

\subsection{Future Prospects}
\label{prosp}
Considering the possible anti-correlation between the P-band fluence
and the L/S-band fluences, there may exist
a pivot frequency between P and L bands.
In the pivot frequency, the fluence variation may be smaller than those in other bands.
Simultaneous observations at a frequency between P and L (for example, $\sim 0.7$ GHz),
and other frequencies may verify the correlation we found.

As for PSR B1133+16, each single pulse, not GRP, can be detected.
\citet{Kr03} claimed that only the pulse at 4.85 GHz showed
a large fluence fluctuation, while
the fluences in other frequencies (0.34, 0.63, 1.41 GHz)
distributed below a few times the mean value.
For the ``GRP-like'' pulses at 4.85 GHz, \citet{Kr03} found
a positive correlation between the flux density at 4.85 GHz
and the spectral index for the frequency range of 0.34--4.85 GHz.
Their correlation is qualitatively similar to 
that which we have found at the L or S band.
Studies based on more spectra of GRPs or normal pulses from other pulsars
are desired to compare with the results of the Crab GRPs.
Since single pulses in a narrow band tend to be affected by the intensity modulation
due to the interstellar scintillation,
ultra-wide-band observations with the next generation telescopes,
such as SKA, will be promising for such studies.

\acknowledgments
First, we appreciate the anonymous referee.
This work is supported by Grants-in-Aid for Scientific
Research Nos. 15J09510 (RM), 15K05069 (TT and KA),
25400227, 16K05291 (KA), 2510447 (SJT), and 16J06773 (SK) from the Ministry
of Education, Culture, Sports, Science and Technology
(MEXT) of Japan.
We appreciate K. Fujisawa, O. Kameya, T. Aoki, K. Niinuma,
and M. Honma for the help in our collaboration.
We also thank N. Hiroshima, Y. Yamakoshi, K. Nagata,
and H. Miyamoto for their helpful cooperation.

\appendix

\section{Flux Density and Error}
\label{est}
For the pulsar data analysis, we adopt the standard coherent dedispersion procedures 
\citep[][]{HR75,LK04} for the data in each frequency band individually.
From the antenna voltage at a certain band after dedispersion $V(t)$,
we average $|V(t)|^2$ for a certain time interval depending on the frequency band
(except for the on-pulse phase) as $\bar {\cal E}$, and similarly obtain its standard deviation $\sigma$.
The S/N at a time bin $i$,
which is defined by the time interval from $t$ to $t+\Delta t$, is written as
\begin{equation}
{\cal S}_i \equiv
\frac{1}{\sigma} \left(
\frac{1}{\Delta t}\int_{t}^{t+\Delta t} |V(t')|^2 dt' - {\bar{\cal E}} \right).
\end{equation}
We generate time-series of the S/N as $i=1$, 2, 3, $\ldots$
for each band.
In our observation, the Crab nebula contributes to the flux.
In this case, we obtain the flux density
as ${\cal F}_i = C {\cal S}_i$,
where the conversion factor $C$ is given by the radiometer equation \citep{Di46, LK04}
as
\begin{equation}
\label{eqn:radiometer}
C=\frac{\mathrm{SEFD}+S_{\rm CN}}{\sqrt{\Delta \nu \Delta t n_{\rm p}}}.
\end{equation}
In equation (\ref{eqn:radiometer}), 
$\Delta \nu$ is the observation bandwidth [Hz],
$n_{\rm p}$ is the number of polarization directions,
$\mathrm{SEFD}$ [Jy] is the system equivalent flux density, 
and
$S_{\rm CN}$ [Jy] is the received flux density of the Crab nebula.
For all the instruments in this paper,
$n_{\rm p}=1$.
The values of $\Delta \nu$, $\mathrm{SEFD}$ and $S_{\rm CN}$ are tabulated
in the fourth to sixth columns of Table \ref{t1}.

For L, S, and C bands, we use the Crab nebula itself as a calibration source.
The method to obtain $S_{\rm CN}$ is as follows.
We assume that the antenna pattern is axisymmetric Gaussian one,
\begin{equation}
P(\theta)=\rm{exp}\left[-4\ln2\left(\frac{\theta}{\theta_{b}}\right)^{2}\right],
\end{equation}
where $P(\theta)$ is the normalized power pattern \citep[][]{Wi13},
and $\theta$ is the distance from the center of the Crab nebula
in the sky coordinate. The half power beam width (HPBW)
$\theta_{b}$ is given by $\lambda/D$,
where $\lambda$ is the observed wavelength and $D$ is the antenna diameter.
From Figure 1 of \citet{Am00}, the spatial intensity distribution
of the Crab nebula is approximated by a linear function of
$\theta$,
\begin{equation}
B_{\rm CN}(\theta)=B_{0} \left(1-\frac{\theta}{\theta_{\rm CN}}\right).
\end{equation}
We adopt $\theta_{\rm CN}=3$ arcmin in all the frequency bands
\citep[e.g.][]{BT15}. The coefficient $B_{0}$ is determined by
\begin{equation}
2 \pi \int_{0}^{\theta_{\rm CN}} d\theta \theta B_{\rm CN}(\theta)=S_{\rm CN,tot}.
\end{equation}
The total flux $S_{\rm CN,tot}$ is calculated from the value in \citet{Ma10} as
\begin{eqnarray}
S_{\rm CN,tot}&=&973\pm19 \left(\frac{\nu_{\rm c}}
{1\mathrm{GHz}}\right)^{-0.296\pm0.006} \nonumber \\
&& \times \exp \left( -\kappa (Y_{\rm obs}-2003) \right)~\mbox{Jy},
\label{Scn}
\end{eqnarray}
where $\kappa=1.67\times10^{-3}/\mbox{yr}$,
$Y_{\rm obs}$ is the epoch of observation in years A. D.,
and $\nu_{\rm c}$ is the center frequency.
Finally, we obtain
\begin{equation}
\label{Scn-rec}
S_{\rm CN}=\frac{\int_{0}^{\theta_{\rm CN}}d\theta \theta B_{\rm CN}(\theta)P(\theta)}
{\int_{0}^{\theta_{\rm CN}}d\theta \theta B_{\rm CN}(\theta)}
S_{\rm CN,tot}.
\end{equation}
A standard on-off observation provides the SEFD as
\begin{eqnarray}
\label{yfac2}
\rm{SEFD}&=&\frac{S_{\rm CN}}{y-1}, \\
y\equiv\frac{P_{\rm on}}{P_{\rm off}}&=&\frac{\rm{SEFD}+S_{\rm CN}}{\rm{SEFD}},
\label{yfac}
\end{eqnarray}
where $P_{\rm on/off}$ is the received power pointing
at the on-source/off-source.
The off-source observation was pointed at a direction $1^{\circ}$ away from the nebula center.

For the X band, we observed Jupiter as a calibration source.
We estimate the SEFD with an on-off observation of Jupiter
assuming that the spatial intensity distribution
of Jupiter is uniform within $\theta < 16.15$ [arcsec].
We use the total flux of Jupiter reported by \citet{Im03}.
Using the value of SEFD estimated above, we estimate $S_{\rm CN}$
with an on-off observation of the Crab nebula.
For the P band, the calibration source was Cyg A.
The flux density of Cyg A is given by \citet{Ba77},
and the receptions of Cyg A and the Crab nebula are set to unity,
because the sizes of those sources are sufficiently smaller than the beam size
for the P band. 

Taking into account the overlaps of the frequency intervals,
the calibrated flux densities for L6--L8 and S1--S5 channels
are synthesized into the values for LH and S bands, respectively.
The statistical error in the flux density ${\cal F}_i$ in each time bin
is $C$.
The systematic error 
$\sigma_{\rm sys}$ includes the uncertainty in the literal flux density
and the time fluctuation of $P_{\rm on/off}$.
For each GRP, setting a time interval with a method described in Appendix \ref{interval},
we integrate the flux to obtain the fluence $F=\Delta t \sum_{i} {\cal F}_i$.
The systematic error in $F$ is written as
\begin{equation}
\sigma_{\rm sys}= a_{\rm sys} F,
\end{equation}
where $a_{\rm sys}$ is defined by using the errors of SEFD
($\Delta \rm{SEFD}$) and the errors of $S_{\rm CN}$ ($\Delta S_{\rm CN}$), as
\begin{equation}
a_{\rm sys}\equiv\frac{\Delta \rm{SEFD}+\Delta S_{\rm CN}}{\rm{SEFD}+S_{\rm CN}}.
\end{equation}
Therefore, the total error for the fluence $F$ is described as
\begin{equation}
\sigma_{\rm tot}=\sqrt{N(C \Delta t)^2+(\sigma_{\rm sys})^{2}},
\end{equation}
where $N$ is the number of time bins in the interval to estimate the fluence. 
The error increases with the number $N$, while the fluence
should converge to a finite value for a significantly large $N$.
When the maximum S/N is below the criterion we define (see the next section)
or $F \leq 3 \sigma_{\rm tot}$, we set the upper limit value as
\begin{equation}
F_{\rm max}=F+\sigma_{\rm tot}.
\end{equation}

\begin{figure}[!htb]
\begin{center}
\includegraphics[width=0.5\textwidth]{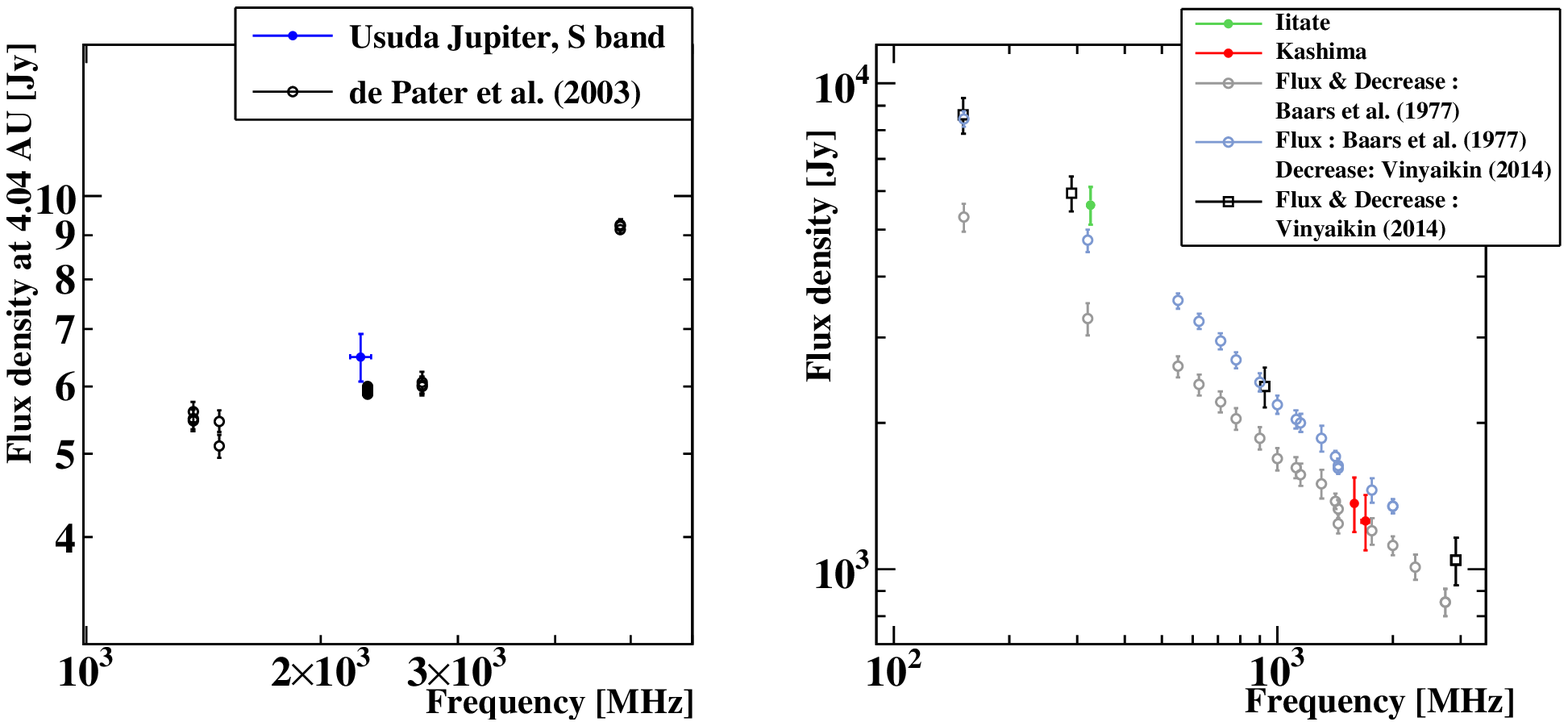}
\caption{Flux densities of Jupiter (left) and Cas A (right).
Our data and previous observations \citep{Im03,Ba77,Vi14} are plotted.
As for Jupiter, our data is normalized
to the flux at the distance of 4.04 au for comparison with the other data.
Taking into account that the fluxes of Cas A show secular decrease
with a frequency-dependent rate,
we plot three cases with different combinations of
the literal flux density and decrease rate
of \citet{Ba77} and \citet{Vi14}. \label{CasCyg}}
\end{center}
\end{figure}

In order to verify our calibration,
we also observed Cas A with Kashima and Iitate telescopes,
and Jupiter at the S band with the Usuda telescope.
We estimated the total flux densities of these objects.
For the intensity distribution of Cas A,
a shell with an outer radius of $130\pm5$ arcsec
and a thickness of $32\pm5$ arcsec are assumed,
based on the observation of \citet[][]{Ro70}. As shown in Figure \ref{CasCyg},
our flux calibration seems to be consistent
with previous observations \citep{Im03,Ba77,Vi14}.

\begin{figure}[!htb]
\epsscale{1.0}
\plotone{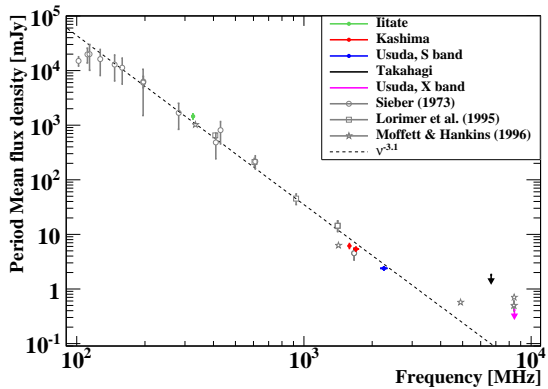}
\figcaption{Period mean flux density of the Crab pulse.
Our data and previous observations \citep{Si73,Lo95,MH96} are plotted.
The dotted line is a reference one showing
$\propto\nu^{-3.1}$ \citep{Lo95}.
For the C and X bands, our observation yields just upper-limit values. \label{NP}}
\end{figure}

We also compare the period mean flux density of the Crab pulse
\citep[e.g.][]{Ma00} with previous studies in each frequency band.
In Figure \ref{NP}, our results agree with previous studies \citep{Si73,Lo95,MH96}.
The significant pulse signals were not detected at C and X bands,
which may be due to the effects of the interstellar scintillation.

\section{Time Interval for Fluence Estimate}
\label{interval}

As long as a GRP is selected in a certain band with the method described
in Section \ref{GRPsel},
the fluences or their upper-limits are estimated for all the bands.
To obtain those values, we need to set a time interval for the integral.
If we adopt a time interval that is too long compared to the actual duration of a GRP,
an excessively large error in $F$ will be obtained.
Each GRP in each observational band shows a different duration.
While the typical duration in LL band is a few $\mu$s,
several GRPs with a duration of $\sim20\ \mu$s were also observed.
Therefore, we determine the time interval for each GRP individually.
A characteristic difficulty in our analysis arises from the very different
duration timescales in the P band and the other L/S bands.
According to the phase distributions of the GRP signals
detected at both the LL and P bands, we empirically established the method
to determine the interval as follows.
According to the method described in Section \ref{GRPsel},
we search for GRPs at P--S bands.
Taking into account the different timescales in P and the other bands,
we divide the GRP samples into two cases:
(1) there are signals above the GRP selection thresholds in the data of
at least one of LL, LH, and S bands.
(2) there are signals above the GRP selection thresholds only in the P band data.

\subsection{GRPs Selected at LL/LH/S Bands}
In the three bands, LL, LH, or S, we occasionally found
a misaligned GRP in different bands or a GRP that has a long quiescent period
between two pulses (see e.g. Figure \ref{sp1}).
Therefore, we use all the light curves in the three bands to determine
the time interval, which is commonly adopted for all the three bands.
The fluences are calculated based on the 1$\mu$s time-bin data.
From the 10$\mu$s time-bin light curves, we first determine a tentative
time interval as follows.
When a GRP is selected in some of the three bands,
at least one time bin of S/N$>6$ exists.
In the three light curves, we search for a single time bin of S/N$>6$
or multiple successive time bins of S/N$>4$.
The time interval $(t_1,t_2)$ is defined as the minimal interval
that contains all the above high S/N time bins.
Then, we generate the time series of the data with 1$\mu$s time-bin
for all the three bands for the time interval
$(t_1-10\mu\mbox{s},t_2+10\mu\mbox{s})$.
From those three data sets, we define the bin numbers at which
the first and last S/N$>3$ signal
appears as $i_{\rm s}$ and $i_{\rm e}$, respectively.
Taking into account the time accuracy between the instruments ($\sim1\ \mu$s),
we integrate the data over time bins from $i_{\rm s}-1$ to $i_{\rm e}+1$
to estimate the fluences or their upper limits for LL, LH, and S bands.

For the P band, the time series of the data is generated with 500$\mu$s time bin.
Since the GRP duration in L/S band is short,
the bin number $i_{\rm m}$ corresponding to the L/S GRP time
is uniquely determined.
When the highest S/N in the interval $(i_{\rm m}-1,i_{\rm m}+2)$
is below four, we set an upper limit of the fluence using the same interval.
This time interval is based on the fact that 95\% of P-band GRPs
have a duration shorter than $2000\ \mu$s.
Including the case that a P band GRP is selected,
when the highest S/N is above four, we set the highest bin as the starting point,
and extend the interval back and forward as long as the successive time bin
has S/N$>2$.
This is because a strong P-band GRP tends to show a long tail
in its light curve.
We integrate over this time interval of S/N$>2$ and estimate the fluence.

\subsection{GRPs Selected at Only P Band}

When a GRP is selected at only P band,
the method to determine the time interval for the P band is the same as
in the previous case.
In this case, the highest signal in the other bands
is below the selection threshold.
In the duration of a P-band GRP, there are $1000$ or more time bins
in the LL, LH, and S band light curves of $\Delta t=1\ \mu$s.
In those large data sets, several time bins of S/N$\sim 3$ or 4 may appear
as the statistical fluctuation.
Therefore, it is dangerous to search for a weak GRP signal
in such a long timescale.
In this case, we set upper limits of the fluences at LL/LH/S bands.
From all the three light curves of $\Delta t=10\ \mu$s,
we search for the highest S/N time bin.
Centering this time bin, we integrate the data over 30$\mu$s
to obtain the upper limits.

\section{Error Estimate for LH and S Bands}
\label{binning1}

\begin{figure}[!htb]
\epsscale{1.2}
\plotone{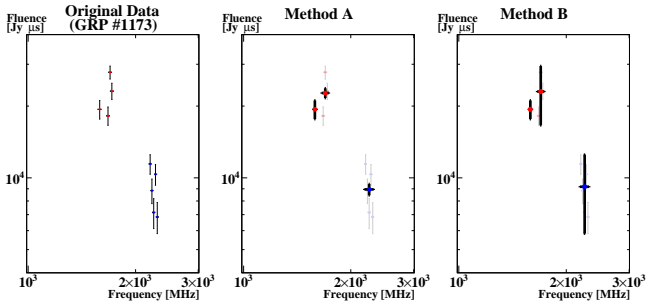}
\figcaption{Error estimates for the fluences at LH and S bands.
Left: original data points of the four channels in the L band (red),
and five channels in the S band (blue) for GRP \#1173.
Middle: synthesized data points and errors at LH and S bands with the
method A.
Right: the same as the middle panel but for the conservative error estimate
with the method B.
\label{bin}}
\end{figure}

In our analysis, we synthesized the flux densities
for L6--L8 and S1--S5 channels into LH and S bands, respectively.
While the statistical errors in the synthesized flux
are estimated with the standard manner taking into account
the overlaps of the frequency intervals,
the systematic error is evaluated by
a sum of the systematic errors of the individual channels.
The obtained errors become significantly small owing
to the channel synthesis as shown in the central panel of Figure \ref{bin}.
Hereafter, this standard method is called ``method A'',
which is the method adopted in the analysis in the main text.

The obtained fluences for the individual channels frequently show
significant variations within their narrow frequency range
as seen in the left panel of Figure \ref{bin}.
As \citet{Ka10} claimed, the fluence fluctuation in the narrow bands
may be intrinsic property in the GRP spectra similarly to
the band spectral structure seen in \citet{HE07}.
Although the modulation due to the diffractive interstellar scintillation
may be negligible at the channels in the L and S bands as discussed in Section \ref{3band},
the fluctuation may be the result of the unknown effects
of the interstellar scintillation.
In order to take into account this uncertainty,
we attempt another conservative method for the error estimate of the synthesized fluences,
``method B''.
In this method, the actual fluence is assumed to be
between the highest and the lowest values in the fluences for the channels
in each band.
The upper and lower limits in the fluence $F_{\rm max}$ and $F_{\rm min}$
are defined as follows,
\begin{eqnarray}
F_{\rm max}\equiv \max (F_{\rm{ch}}+\sigma_{\rm{ch}}),
\quad
F_{\rm min }\equiv \min (F_{\rm{ch}}-\sigma_{\rm{ch}}), \nonumber \\
\end{eqnarray}
where $F_{\rm{ch}}$ and $\sigma_{\rm{ch}}$
are the fluence and its $1\sigma$ error, respectively,
at channel $\rm{ch} = \rm{L6, L7, L8}$ or S1, S2, S3, S4, S5.
In the method B, the representative fluence $F$
and its $1\sigma$ error $\sigma_{\rm tot}$ are assumed as
\begin{eqnarray}
F&=&\frac{F_{\rm max}+F_{\rm min}}{2}, \\
\sigma_{\rm tot}&=&F-F_{\rm min}=F_{\rm max }-F.
\end{eqnarray}
As shown in the right panel of Figure \ref{bin}, the method B leads to larger errors.
When there is a channel with a fluence below $3 \sigma_{\rm{ch}}$,
we set an upper-limit $F_{\rm max}$ on the fluence.
In this conservative method, even if an S/N value for the synthesized channel
is above the selection threshold, there are cases
in which just the upper limit is set on the synthesized fluence.
Therefore, some fraction of GRPs detected with the method A are judged as non-detection.
The sample number in the method B becomes slightly smaller than that in the method A
(compare Tables \ref{tcatMP} and \ref{tcatMP-binning1}).

When we use the data generated with the method B in the spectral analysis,
95.6\% of MPGRPs and 97.0\% of IPGRPs are judged to be consistent
with SPLs at a significance level of 5\%
(see Table \ref{tcatMP-binning1}).

\begin{table*}
\centering
\scriptsize
\caption{The same table as Table \ref{tcatMP},
but with the conservative error estimate, method B. }
\label{tcatMP-binning1}
\begin{tabular}{rrrrrrr}
\hline
 & & \multicolumn{2}{c}{SPL Hypothesis (MPGRP)}  & & \multicolumn{2}{c}{SPL Hypothesis (IPGRP)}\\
Detection Y/N & Group$^{a}$ & Consistent  & Inconsistent && Consistent  & Inconsistent\\
(P, LL, LH, S band)& &($\hat{\chi}_{\mbox{min}}^{2}<5.99$)    & ($\hat{\chi}_{\mbox{min}}^{2}>5.99$) && ($\hat{\chi}_{\mbox{min}}^{2}<5.99$)    & ($\hat{\chi}_{\mbox{min}}^{2}>5.99$)\\
\hline\hline
(Y, Y, Y, Y)                     & I & 163 & 20 && 26 & 2 \\
(Y, Y, Y, N)                             & I & 61 & 3 && 12 & 0 \\
(Y, N, N, Y)                                           & I& 3 & 1 && 0&0 \\
(Y, Y, N, Y)                                 & I& 9 & 0 && 0 & 0 \\
(Y, N, Y, Y)                                & I & 1 & 0 && 0 & 0 \\
(Y, Y, N, N)                                         & I & 15 & 0 && 6 & 0 \\
(Y, N, Y, N)                                        & I & 3 & 0 && 1 & 0 \\
(N, Y, Y, Y)                             & II & 944 & 59 ($\alpha>2$ : 1) && 60 & 0 \\
(N, Y, N, Y)                                        &II & 165 & 1 && 10 & 0 \\
(N, N, Y, Y)                                        &II & 39 & 2 && 1 & 0 \\
(N, Y, Y, N)                                     &\nodata & 408 & 25 && 28 & 2 \\
(N, N, N, Y)                                      & \nodata& 146 & 1 ($\alpha>2$ : 1) && 4 & 0 \\
(N, Y, N, N)                                                  & \nodata& 278 & 15 && 23 & 2 \\
(N, N, Y, N)                                                 & \nodata& 79 & 2 && 2 & 0 \\
(Y, N, N, N)                                       & III & 480 & 1 ($\alpha<-5$ : 1) && 53 & 1 ($\alpha<-5$ : 1)\\
\cline{2-7} \\
Total                                                  &  & 2794 & 130  && 226 & 7 \\
                                                          &  &  &       ($\alpha>2$ : 2, $\alpha<-5$ : 1) && & ($\alpha>2$ : 0, $\alpha<-5$ : 1)\\
\hline
\end{tabular}
\\ {\scriptsize a: We categorize the GRPs into three groups, from I to III. See Section \ref{cate}.}
\end{table*}

\section{Monte Carlo Simulations for the apparent correlations}
\label{montecarlo}

Some correlations between the fluences and spectral indices are found in our analysis.
In order to probe the intrinsic distributions of the fluences and indices,
we simulate multi-frequency observations of SPL GRPs with the Monte Carlo method,
assuming that an intrinsic correlation between the fluence and hardness is absent.
However, we should notice that for given fluence
at L/S band, a smaller spectral index implies
a larger energy release at the P band, where most of the energy is released.
Namely, the uncorrelation between the fluences at L/S band and $\alpha$
is equivalent to the assumption of
an anti-correlation between the total energy release
and hardness.
As will be shown, this assumption seems most likely to reproduce
the observed correlations.

\begin{figure}[!htb]
\epsscale{0.8}
\plotone{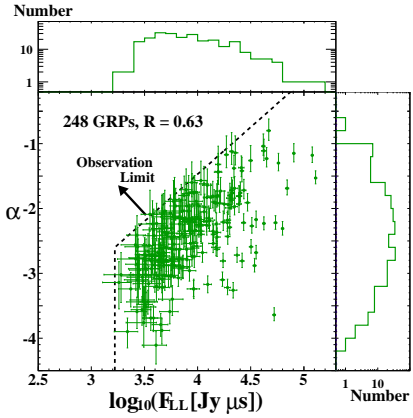}
\figcaption{Scatter plot of $F_{\rm LL}$ and $\alpha$ for the Group (I) MPGRPs selected with the method B. \label{real}}
\end{figure}

In Figure \ref{real}, we show the scatter plot of Group (I) MPGRPs detected at the LL band,
obtained with the method B (see Appendix \ref{binning1}),
for the fluence at the LL band ($F_{\rm LL}$) and the spectral index ($\alpha$).
While the sample number is larger than that with method A in Section \ref{corLS},
the similar correlation is seen.
To reproduce this correlation, we generate pseudo GRP samples whose fluence distribution at the LL band
follows the power law with index -2.98 as observed (see Table \ref{tFluenceIndex})
above $1260$Jy~$\mu$s.
The suppression in the lower fluence region (``roll-over'' feature) seen in the distributions
shown in Figure \ref{fig:fluenceDistribution} may be due to the decline of the detection efficiency.
The detection efficiency in the lower fluence region for each band is determined to reproduce
the roll-over shape.
For each pseudo GRP, we obtain $\alpha$ following a distribution assumed in advance,
independently of $F_{\rm LL}$.
The fluence at other bands is calculated with the obtained $F_{\rm LL}$ and $\alpha$.
We judge the detection of the pseudo GRPs taking into account the detection efficiency at each frequency band,
and then categorize the GRPs into the groups described in Section \ref{cate}. 
The GRP sample generation continues until the number of GRPs detected at both P and LL bands reaches 271,
which is the number of GRPs detected at both the P and LL bands in our observation samples.
Since 248 GRPs (samples in Figure \ref{real}) are judged to be consistent with a SPL,
we randomly extract 248 GRPs from the 271 pseudo samples.
We perform such trials $10^{4}$ times for a certain distribution of $\alpha$.

\begin{figure}[!htb]
\epsscale{1.15}
\plottwo{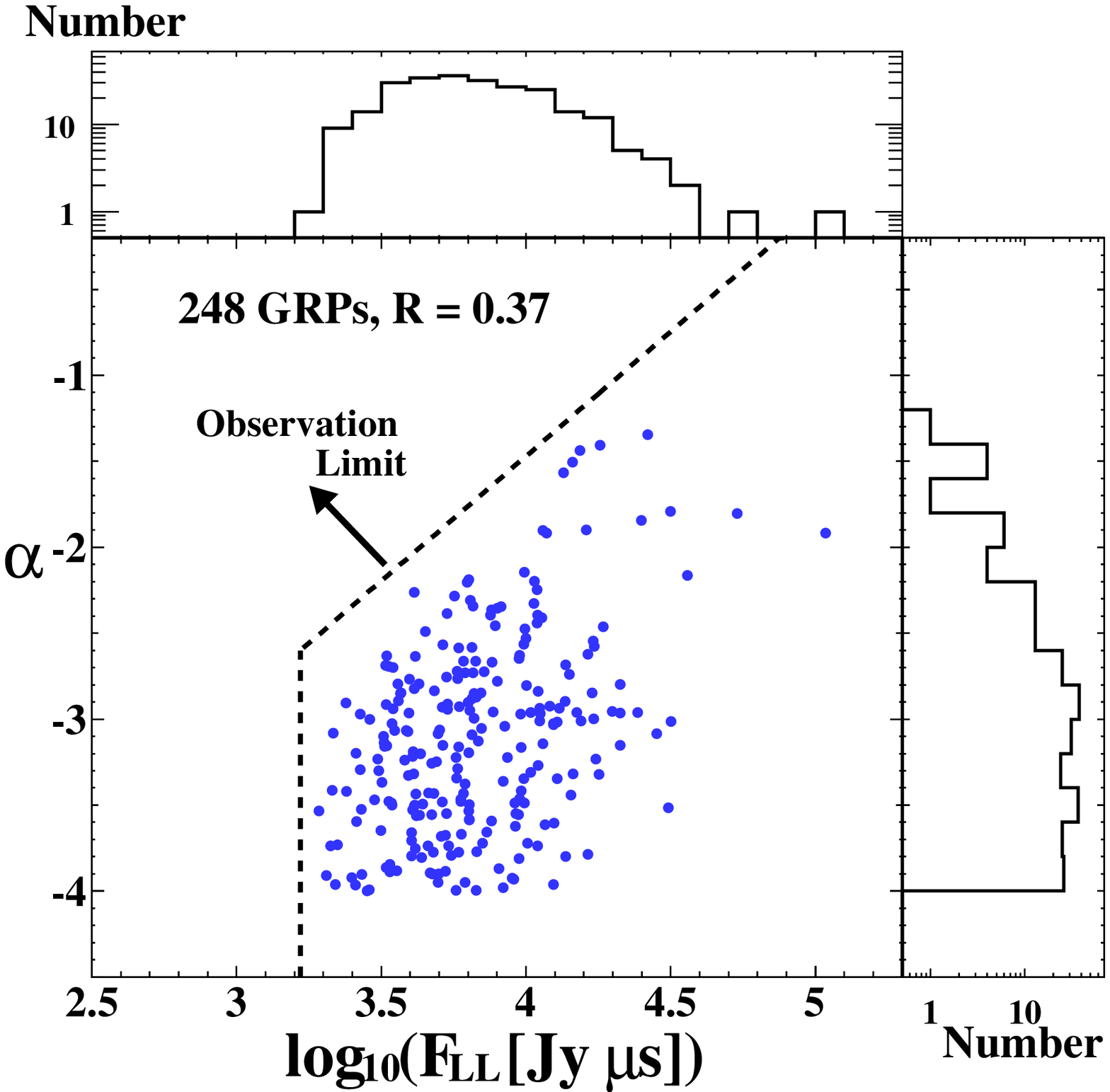}{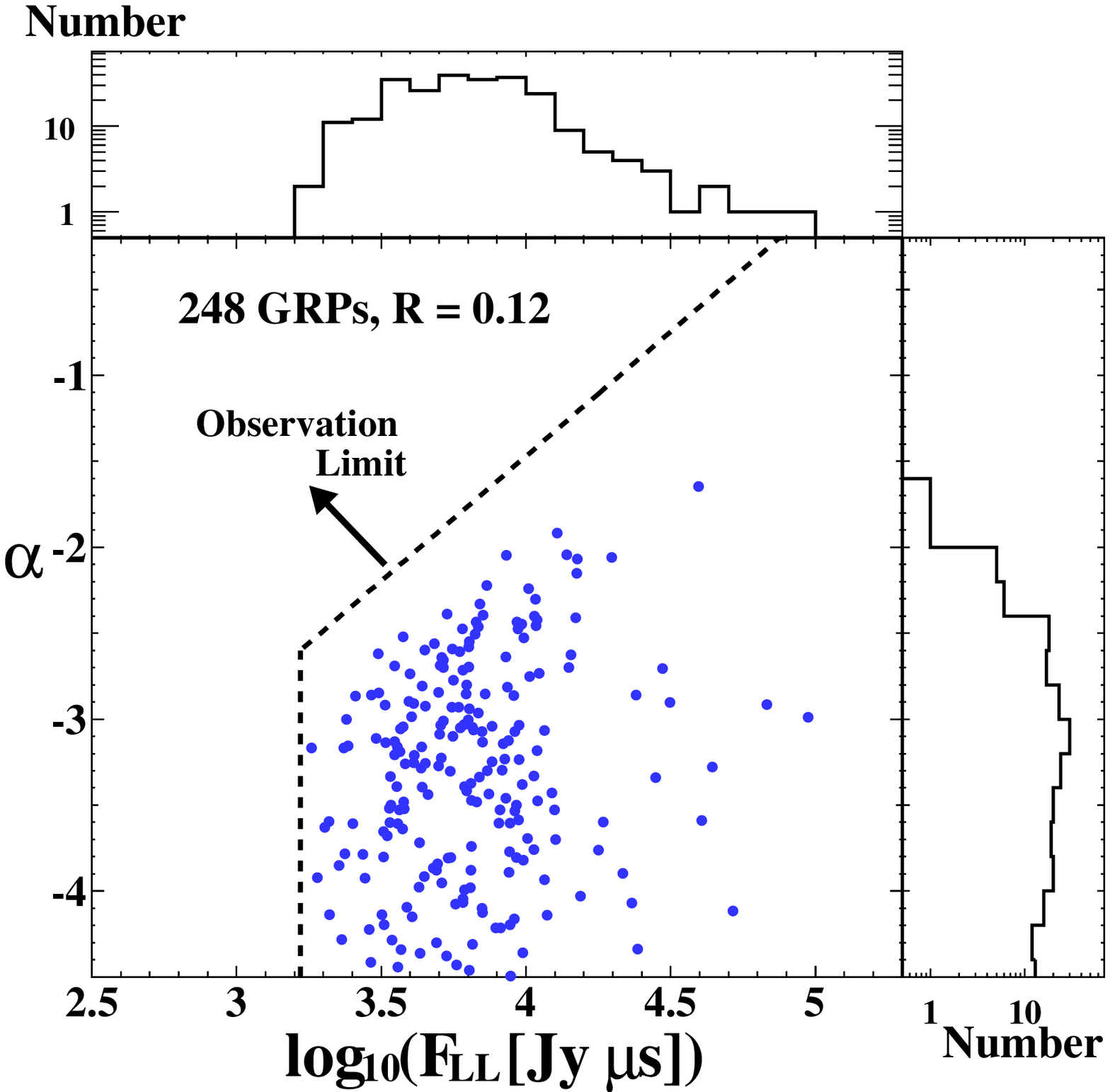}
\figcaption{Scatter plots of $F_{\rm LL}$ and $\alpha$ for pseudo samples with a uniform $\alpha$ distribution from $-4.0$ to $0.0$ (left), 
and a Gaussian one with $(\bar{\alpha},\sigma_\alpha)=(-3.0,1.0)$ (right). \label{alphagaus}}
\end{figure}

First, we test a uniform distribution from $\alpha=-4.0$ to $0.0$,
and a Gaussian distribution with the mean value of $\bar{\alpha}=-3.0$,
which is close to that for the normal pulse \citep[e.g.][]{MH99}, and the standard deviation $\sigma_\alpha=1.0$.
For those models, examples of the scatter plots obtained from one of the trials
are shown in Figure \ref{alphagaus}.
Apparently, the distributions seem different from the observed one.
We calculate the Spearman rank correlation coefficient $R_{\rm LL}$ for those 248 pseudo
samples in each trial,
but none of the trials reaches $0.63$, the value for the actual Group (I) GRPs (see Figure \ref{Rhist}).
The uniform distribution and $\bar{\alpha}=-3.0$ are excluded.

\begin{figure}[!htb]
\epsscale{1.15}
\plottwo{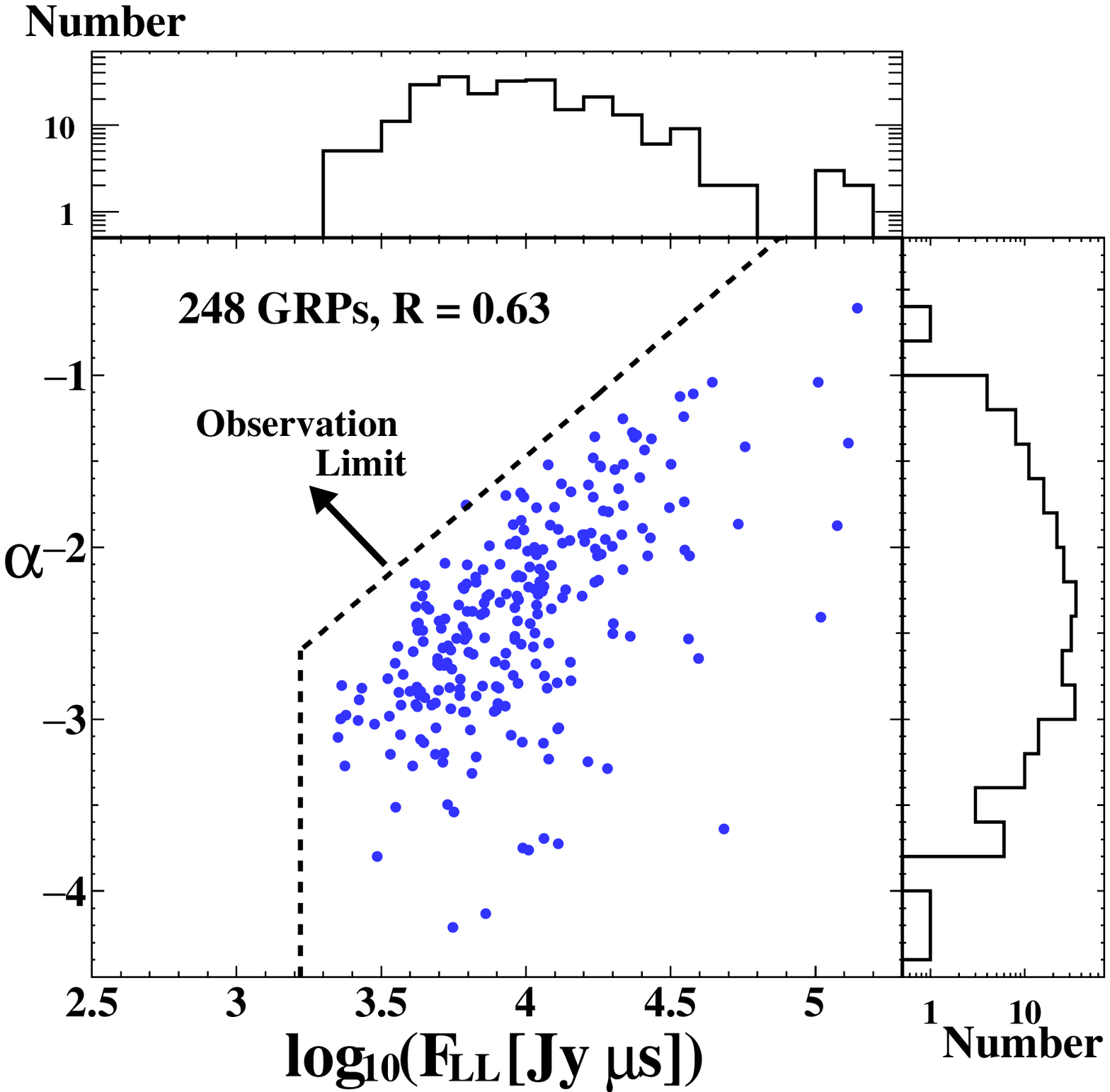}{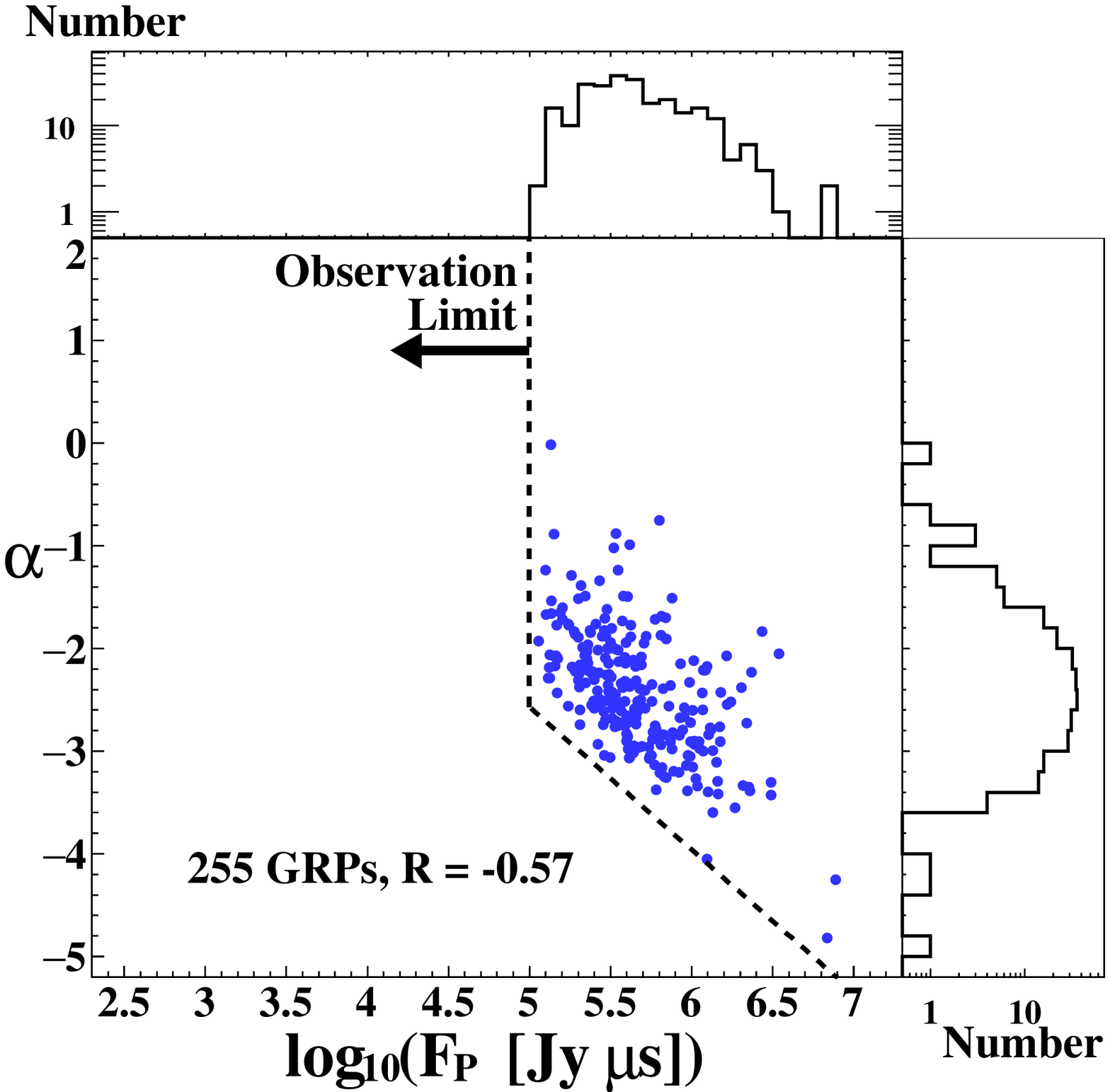}
\figcaption{Scatter plots for pseudo samples with a Gaussian distribution for $\alpha$
with $(\bar{\alpha},\sigma_\alpha)=(-1.0,0.9)$ in the LL (left) and P (right) bands. \label{alphagaus_Rhigh}}
\end{figure}

\begin{figure}[!htb]
\epsscale{1.15}
\plottwo{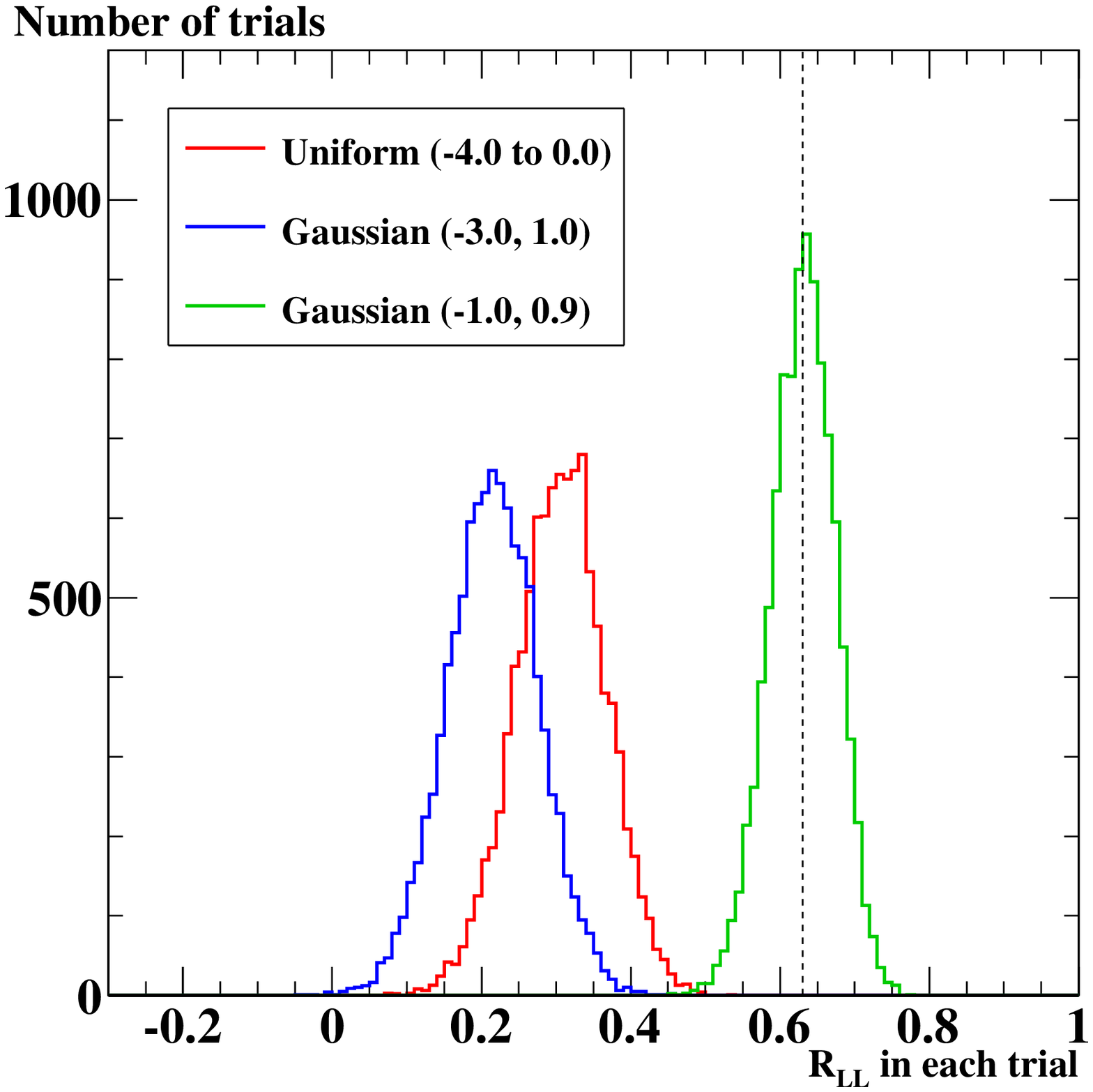}{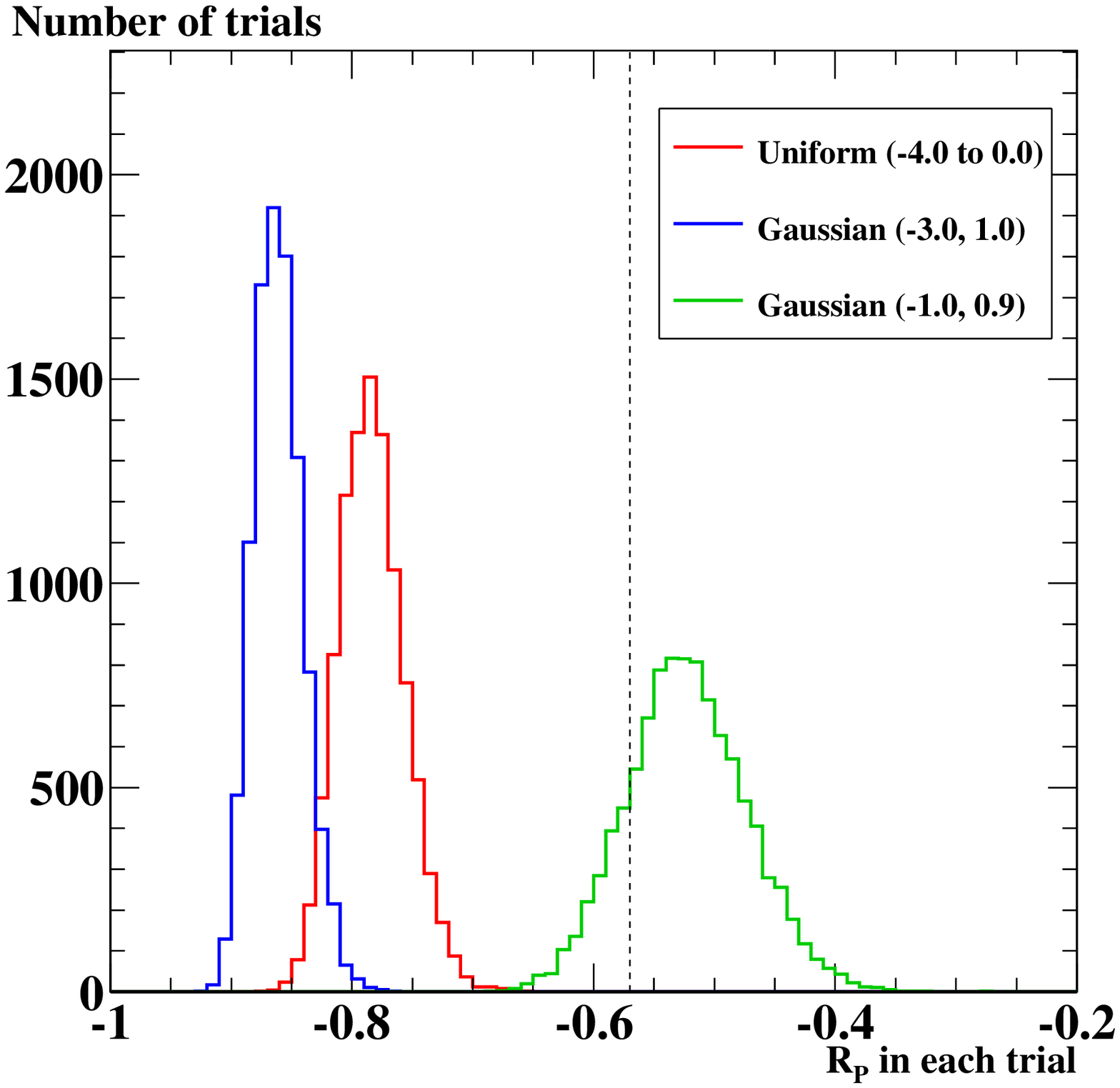}
\figcaption{Distributions of the Spearman rank correlation coefficient in the LL (left) and P (right) bands
for the pseudo samples of the different $\alpha$ distributions.
In each model, we test $10^{4}$ trials. 
The dashed lines represent the values obtained from the real data with the method B. \label{Rhist}}
\end{figure}

However, if the typical index is harder than that for the normal pulse,
only a small fraction of dim GRPs at the LL band is soft enough to be detected at the P band.
Here, we show the results for the Gaussian distributions with
$(\bar{\alpha},\sigma_\alpha)=(-1.0,0.9)$ as an example.
The distributions in the LL and P bands for those models, as shown in Figure \ref{alphagaus_Rhigh},
are similar to Figures \ref{real} and \ref{flualpha}, respectively.

\begin{figure}[!htb]
\epsscale{1.15}
\plottwo{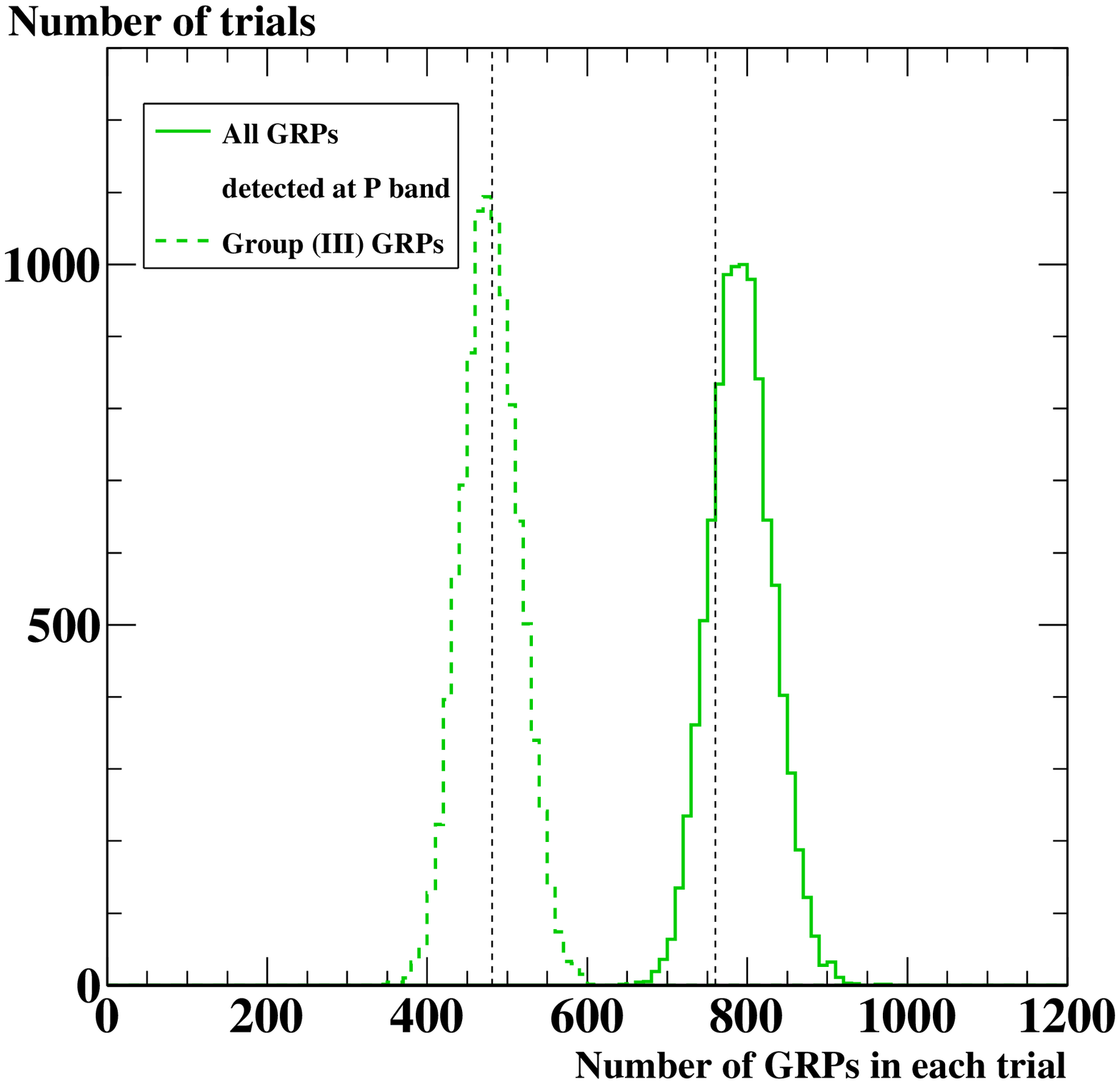}{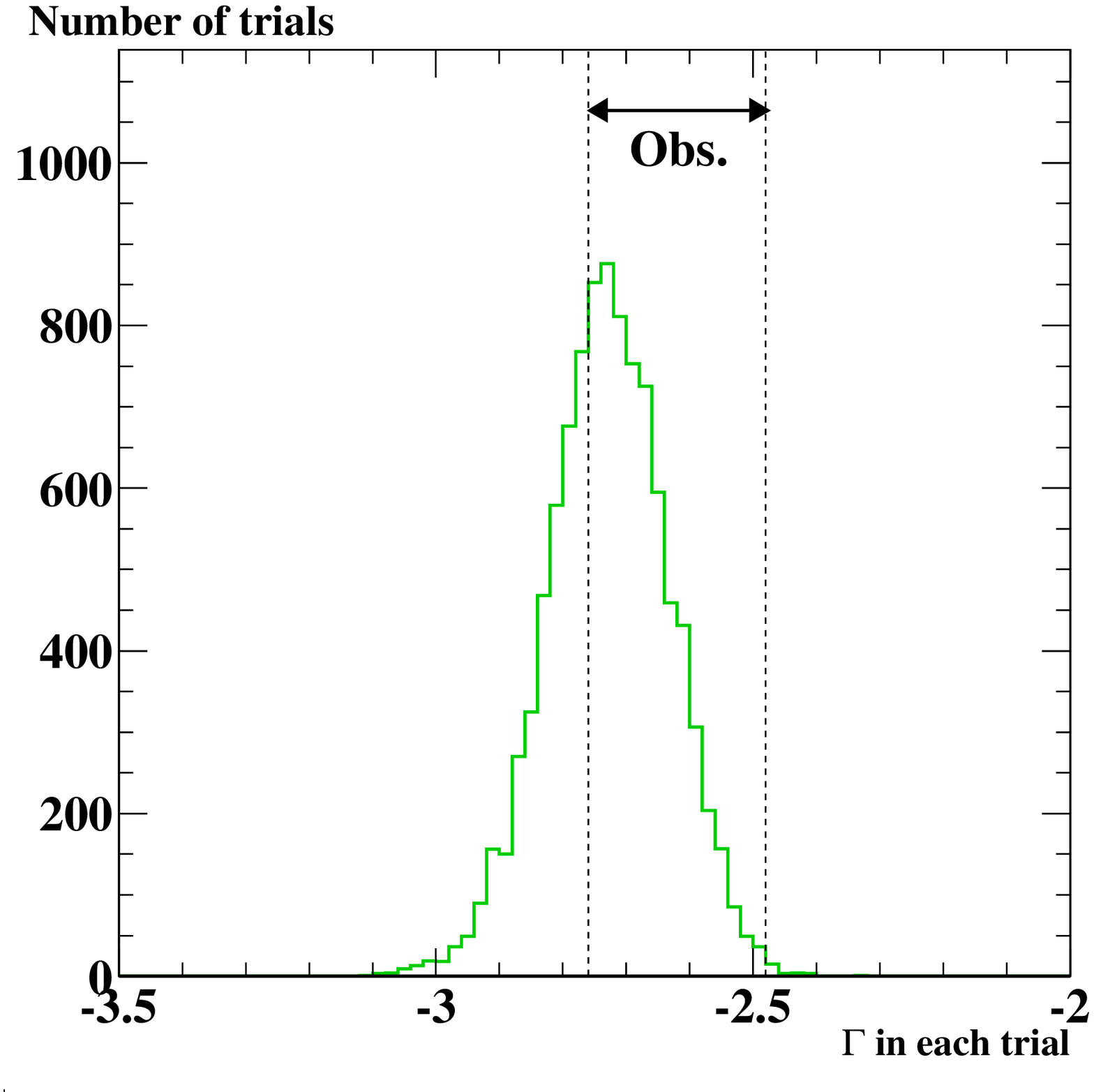}
\figcaption{Distributions of the number of GRPs detected at P band,
the number of Group (III) GRPs (left), and
the power-law index in the fluence distribution at P band (right)
for the pseudo samples of the Gaussian $\alpha$ distribution of $(\bar{\alpha},\sigma_\alpha)=(-1.0,0.9)$,
based on $10^{4}$ trials.
The dashed lines represent the values obtained from the real data.
\label{NPhist}}
\end{figure}

Actually, in this model, the distributions of the Spearman rank correlation coefficient
is consistent with the obtained values from our observation as shown in Figure \ref{Rhist}.
Moreover, the number of GRPs detected at the P band, the number of Group (III) GRPs,
and the power-law index of the fluence distribution at the P band in this model roughly agree
with our observation
(see Figure \ref{NPhist}).
This model overpredicts the number of the GRPs detected at only the S band.
Many hard GRPs ($\alpha>0.0$) contribute to those S-band GRPs.
This may indicate that the $\alpha$ distribution is asymmetric.
When we artificially suppress GRPs with $\alpha \gtrsim 0.0$ compared to the exact Gaussian distribution,
the correlations are maintained,
and the number of GRPs detected at only S band becomes consistent with our real samples.
The actual distribution may have a sharp peak around $-1$ and one-sided long tail
at $\alpha<-1$.

\end{document}